\documentclass{elsarticle}

\usepackage[english]{babel}

\usepackage[paperheight=11in,
   paperwidth=8.5in,
   top=1in,
   bottom=1in,
   left=1in,
   right=1in]{geometry}

\usepackage{amsmath,amssymb}
\usepackage{amsthm}
\usepackage{graphicx}
\usepackage{tikz}
\usepackage{tikz-3dplot}
\tikzset{fontscale/.style = {font=\relsize{#1}}}
\usetikzlibrary{positioning, shapes}
\usepackage{color}
\usepackage{float}
\usepackage{caption}
\usepackage{subcaption}
\usepackage{enumerate}
\usepackage[colorlinks=true, allcolors=blue]{hyperref}
\usepackage{wrapfig}

\newtheorem{definition}{Definition}

\newcommand{\R}{\mathbb{R}}

\begin{document}

\begin{frontmatter}
\journal{Communications in Nonlinear Science and Numerical Simulation} 
\title{
Denoising 3D images: robustness of persistent homology measures}

\author[a]{Ebru Dagdelen}
\ead{<end@njit.edu>}
\author[a]{Aakash Karlekar}
\ead{<ak2869@njit.edu>}
\author[a]{Manav Arora}
\ead{<ma2839@njit.edu>}
\author[a]{Matthew Illingsworth}
\ead{<mri8@njit.edu>}
\author[a]{Jonathan Jaquette \corref{cor1}}
\ead{<jonathan.jaquette@njit.edu>}
\author[a]{Linda J.\ Cummings}
\ead{<linda.cummings@njit.edu>}
\author[a]{Lou Kondic}
\ead{<kondic@njit.edu>}

\address[a]{Department of Mathematical Sciences, New Jersey Institute of Technology,
Newark, NJ 07102, USA}

\cortext[cor1]{Corresponding author.}

\pdfstringdefDisableCommands{%
  \def\corref#1{}%
  \def\cnotenum{}%
  \def\cortext#1#2{}%
}

\begin{abstract}

  When computing sub/super-level-set  persistent homology (PH), the effect of noise may introduce millions of (short-lived) topological generators, presenting an obstacle to both the computation of PH of large 3D images, and any analysis of PH that incorporates the number of generators. As such, it is often necessary to denoise the data before computing its PH. 
  We analyze the PH of synthetic 3D images of porous media in the presence of spatially uncorrelated noise, and perform a comparative analysis of various topological measures (e.g. bottleneck distance, Wasserstein distance, persistence statistics and persistence images) to assess their robustness to both noise and the denoising process (i.e. adding spatially uncorrelated Gaussian noise, and denoising by either a Gaussian convolution or a machine learning approach). 
\end{abstract}


\end{frontmatter}

%

\section{Introduction} \label{Introduction}

Topological Data Analysis (TDA)~\cite{Porter2023pt} and in particular persistent homology (PH) can be successfully applied across a wide range of scenarios to effectively analyze and measure topological properties of a material or biological sample based on two-dimensional (2D) or three-dimensional (3D) images. In this context PH finds application across many fields, for example: it is used in medical settings for brain networks~\cite{Aktas2019}, bone morphology~\cite{pritchard2023persistent}, and tumor shape~\cite{Somasundaram2021} analyses; in physics to analyze composite materials~\cite{MUSO2026107721} and to improve understanding of the cosmic structure of regions of space~\cite{Wilding2021}; and in computational settings for image, network, and clustering analyses~\cite{Otter2017}. Depending on the specific information one wants to extract from an image, the presence of experimental noise (which may be pixel/voxel independent and spatially uncorrelated), even at low levels, may give rise to significant errors in PH-based predictions. While image denoising methods are available \cite{jain2016survey,fan2019brief}, it is often unclear how best to tune denoising parameters to remove noise without losing the signal (here, the desired topological information). It is therefore important to be able to quantify the robustness of key topological measures, both to added noise, and to the chosen denoising procedure. 

In this paper we focus on the specific case of three-dimensional (3D) digitized images of porous media, relevant in a wide range of natural and industrial processes such as environmental remediation, secondary
and tertiary oil recovery and CO$_2$ sequestration, see, e.g.,~\cite{Neumann2021,martinez_water24}. Such images, which can be obtained from a 3D scan of a sample from a porous material \cite{blunt-awr-2013, kia_2023, sundaramoorthi-iecr-2016} or be synthetically generated \cite{Mosser2017, zhou2022novel, zhou2023computational, torquato2002random}, form a rigorous testbed for denoising methodology, as porous media have intricate multiscale topological features such as interconnected pores and channels that make their analysis particularly challenging. These features strongly influence how the media function in applications, particularly those involving fluid flow. Thus, accurately characterizing the internal pore structure of these materials and quantifying the heterogeneity is essential for understanding (and possibly controlling or optimizing) such processes, an idea that emerges strongly from Gerritsen \& Durlofsky's review article on fluid flow through oil reservoirs~\cite{gerritsen-2005-arfm}. PH provides us with an effective way to do this~\cite{Otter2017, Rocks2020RevealingHomology, carlsson, Edelsbrunner_Letscher_Zomorodian_2002, edelsbrunner2008persistent}, but to obtain reliable results one must be able to control for the presence of experimental noise in images, which may arise from many different sources such as inherent measuring noise, thermal noise, or other application-dependent effects. An important factor here is the characterization of the noise; we assume that the multiple sources can be represented by i.i.d. random variables, allowing us to model the resultant combined noise as normally distributed by the Central Limit Theorem. 

In the context of analyzing porous media, digital representations are typically either binary or grayscale. In a binary representation, each voxel in the image can take on one of two values, representing either void space or impermeable material, while a grayscale representation approximates a material in which there is continuous variation of density or permeability, captured by assigning a grayscale value to each voxel. In the present work, we start with 3D grayscale images of porous media. Through a thresholding process based on the grayscale level of each voxel in the image, we construct a nested sequence of geometric layers called a topological filtration. Specifically, at a given grayscale threshold level, the subset of the image comprising the voxels with grayscale value brighter than or equal to the threshold constitutes the current geometric layer. We can study the topological properties of each of these filtration layers by computing its relevant homology groups. These groups are most simply described in terms of Betti numbers, of which there are three for a 3D sample, $\beta_0,\beta_1,\beta_2$, corresponding to the numbers of connected components, handles, and cavities found within each layer, respectively~\cite{munkres-pp-1997, edelsbrunner2008persistent}. PH looks to synthesize this information by quantifying the topological features that persist across many layers of a filtration. As we progress through a filtration, the threshold value at which a feature appears is referred to as its birth number. Similarly, when we encounter a threshold value at which a feature disappears, we refer to it as the feature's death number. We can then visualize the persistence of features as points (topological generators) in a 2D graph, where one axis represents each feature's birth number and the other its death number. By plotting the corresponding birth-death pairs for the features we encounter in our filtration, we obtain a Persistence Diagram (PD) for a given homology group. We can use a PD to generate Persistence Barcodes, Persistence Landscapes, and Persistence Images, all of which represent various aspects of the topological composition of our porous material. We say more about each of these representations and their uses in Sec.~\ref{Definitions}.

Before proceeding to analyze representative 3D image data, however, we briefly discuss how the presence of experimental noise impacts PH. Noise presents two main challenges: first, it modifies the results; second, it may make computations of the topological structure extremely intensive. For the type of data of interest here -- images of rock samples -- the issues of noise are compounded by the datasets' topological complexity due to the intricate porous structure. This combination of factors can make it difficult to efficiently extract reliable and meaningful insights from the data using PH: Turkeš \textit{et al.}~\cite{10.1371/journal.pone.0257215} found that the introduction of static noise can weaken the discriminative power of PH outputs such as persistence diagrams, landscapes, and images. Intuitively, this is because even a small amount of voxel-wise noise introduces a large number of short-lived topological features, appearing as a large number of points near the diagonal in the PD, whose number is essentially proportional to the number of voxels in the image. While the bottleneck stability theorem \cite{StablityPersistenceDiagrams} guarantees that image noise  will perturb the bottleneck distance of the original image by at most the maximum size of the noise, the various vectorizations of PDs do not have such robust stability guarantees.  Statistical vectorization (which encodes statistical information, such as the number of PH generators, statistics of their birth/death times, etc.) is particularly vulnerable to the effects of pixel-independent noise. To touch on the computational challenge, we remark that efficient methods to compute the PH of grayscale images have been developed based on discrete Morse theory \cite{robins2011theory,mischaikow2013morse} and the state-of-the-art methods are capable of processing 3D grayscale images of size at most $2048^3$ voxels \cite{wagner2023slice}. These methods are very efficient at reducing the computational cost in a typical case, however, the massive addition of generators coming  from voxel-independent noise results in  a worst-case run-time scenario. 

To address these challenges, one must develop methods that reduce computational complexity and mitigate noise of a given dataset, while still preserving its key topological features. Image denoising is well studied, with a rich literature \cite{jain2016survey,fan2019brief}, and recent progress has been made in preserving topological information \cite{al2024cubical,TopoRSNet} under denoising. For our purposes, we aim to produce a denoised image whose PH is both computationally less demanding than the noisy one, and acceptably close to that of the true image. Effective resolution of this challenge is an active area of ongoing research. For example, Henneuse ~\cite{henneuse2024persistence,henneuse2026persistence} sought to estimate persistence information directly from noisy or incomplete signal data through the use of plug-in estimators, image persistence, and the minimax rate. Others have investigated methods to denoise the noisy image data prior to analyzing it; Tang \textit{et al.}~\cite{TopoRSNet} proposed a machine learning (ML) based approach to adjust the resolution of micro-computed tomography ($\mu$CT) images in a manner that preserves their representative features. Recent literature is comprehensively discussed by Tawfik \textit{et al.}~\cite{10.3389/frwa.2021.800369}, who conducted a physics-based comparative study on how various denoising methods, ranging from Gaussian filters to Deep Learning approaches, perform when denoising samples of porous media images. 
  
In the present work, we study the robustness of various topological measures of an image to the noising/denoising process. We work with selected noisy image data representative of porous media, and apply a standard denoising method using Gaussian convolution with a smoothing parameter $\sigma$ to smooth the data and reduce noise. We additionally consider a machine learning approach, training a model to learn a non-linear mapping between the noisy and original images without the manual parameter tuning that Gaussian smoothing requires. Ideally, for a robust topological measure, we will largely remove noise from our data rather than essential information, and with minimal tuning of the denoising parameters. 
 
\begin{figure}[h!tbp]
    \resizebox{\textwidth}{!}{
        \centering
        \begin{tikzpicture}[node distance=2cm]
            \node[draw, thick, rectangle, minimum size=2cm, align=center] (box1) {\large Original \\ \large dataset};
            \node[draw, thick, rectangle, minimum size=2cm, below=of box1, align=center] (box2) {\large Noisy \\ \large dataset};
            \node[draw, thick, rectangle, minimum size=2cm, right=of box2, align=center] (box3) {\large Denoised \\ \large dataset};
            \node[draw, thick, trapezium, trapezium angle=70, minimum size=1.333cm, right=of box3, shape border rotate=270] (trap1) {\large PH};
            \node[draw, thick, trapezium, trapezium angle=70, minimum size=1.333cm, above=of trap1, shape border rotate=270, yshift=.666cm] (trap2) {\large PH};
            \node[draw, thick, rectangle, minimum size=2cm, right=of trap1, align=center](box4) {\large Vectorization};
            \node[draw, thick, rectangle, minimum size=2cm, right=of trap2, align=center](box5) {\large Vectorization};
        
            \draw[->, thick] (box1) -- (box2);
            \draw[->, thick] (box2) -- (box3);
            \draw[->, thick] (box1.east) |- (trap2.west);
            \draw[->, thick] (box3.east) |- (trap1);
            \draw[->, thick] (trap1.east) -- (box4);
            \draw[->, thick] (trap2.east) -- (box5);
            \draw[<->, thick] (trap1.north) -- (trap2.south) node[right, pos=0.5] {\large Compare};
            \draw[<->, thick] (box4.north) -- (box5.south) node[right, pos=0.5] {\large Compare};
        \end{tikzpicture}
    }
    \caption{\label{fig:Testing Process} Flowchart of methodology. We start with the original synthetic dataset; add noise to produce a noisy dataset; then apply smoothing, with denoising parameter $\sigma$, to produce a denoised dataset. We then compare the PH of the original and denoised datasets, using metrics such as the bottleneck distance and Wasserstein distance between their PDs, and the distance between vectorizations (such as the persistence landscapes and persistence images) of their PDs.  }
\end{figure}
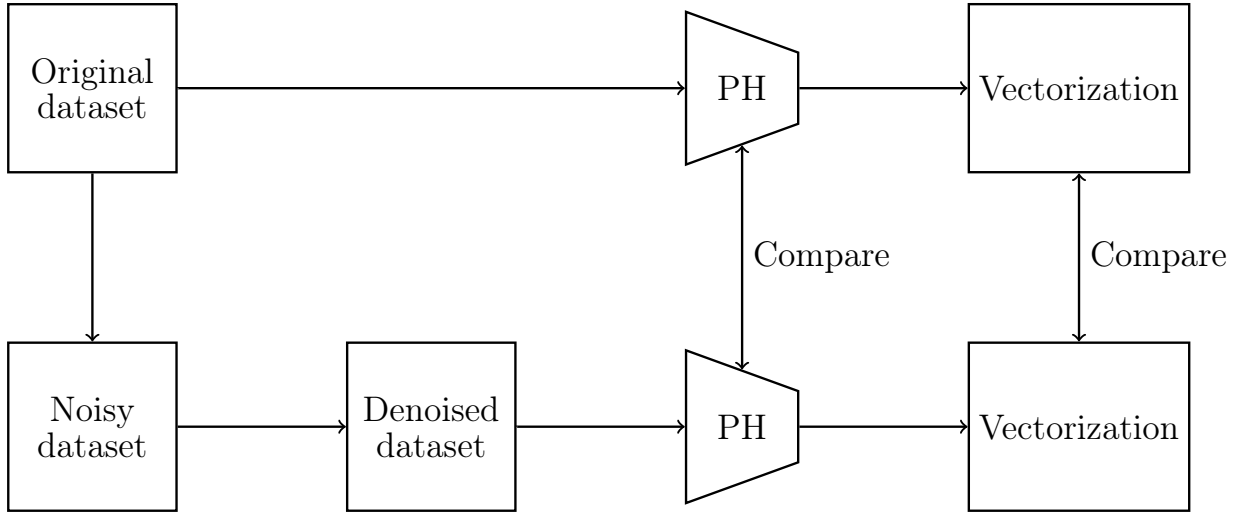

This testing process is schematized in Fig.~\ref{fig:Testing Process}, where we start with a simulated dataset intended to represent a 3D image of a porous medium, which we refer to as the original dataset. We then add controlled static noise from a known distribution to this dataset to produce a noisy dataset. Next, we apply our smoothing method on the noisy dataset for various values of the smoothing parameter $\sigma$, referring to the resulting dataset after smoothing as the denoised dataset.  We identify several topological measures that we use to quantify the difference between the original and denoised datasets. These include the normalized bottleneck and Wasserstein distances between PDs, as well as normalized distances between various vectorizations of the PH, such as Persistent Landscapes and Persistent Images. A robust measure should satisfy two criteria: it should not be unduly sensitive to the level of denoising, and it should remain close to its optimal value over a reasonable range of the denoising parameter $\sigma$. 
Since we know the original state of each dataset before noise was introduced, we can quantify these differences directly and use them to evaluate our denoising algorithm.

We apply our testing process on several datasets, described in Sec.~\ref{Datasets}.  We introduce key topological concepts in Sec.~\ref{Definitions}, discuss our methodology in Sec.~\ref{sec:methodology} and analyze the efficacy of our denoising method through the lens of several measures, which we describe in detail. 
Our results are presented in Sec.~\ref{Results}, and demonstrate that our method is able to minimize acceptably the overall difference between the denoised and original versions of each of our datasets, for some ideal value of the denoising parameter. We also find that, in certain scenarios, some of our measures are more robust than others. We conclude in Sec.~\ref{sec:discussion} with a discussion and summary of the key findings.

\section{Datasets} \label{Datasets}
  In this section, we provide details on the specific datasets we use to test our denoising scheme, including how they are generated, their relevant properties, and how they compare to actual experimental samples of porous media. These are, in summary, a synthetic 3D image generated using Fourier series, which we refer to as the \textbf{Fourier dataset}; a dataset based on random, overlapping spheres generated using the PuMA library, which we refer to as the \textbf{PuMA dataset} \cite{puma2021}; and a Worley noise based dataset generated using the PyFastNoiseSIMD python wrapper which we refer to as the \textbf{Cellular dataset} \cite{pyfastnoisesimd}. In general, the images considered are grayscale 3D cubes, containing $N_v^3$ voxels, where each voxel is associated with an integer grayscale value between 0 and 255.  
  For the Gaussian denoising approach we set $N_v=254$, but to keep computing time manageable
we use a smaller domain size of $N_v=128$ when employing machine learning to denoise the images. 

For each dataset type (described below), we generate 10 independent realizations using different random seeds. All topological measures reported in Sec.~\ref{Results} represent averages over these 10 realizations.
 
\subsection{Fourier Dataset}

We first define the periodic function $F:[0,2\pi]^3 \to \mathbb{R}$ by 
\begin{align*}
    F(x, y, z) &= \sum_{i,j,k=0}^{M}  
    \frac{1}{1 + i^{1/2} + j^{1/2} + k^{1/2}}
    \Big( 
    a_{ijk}^{(1)} \cos( i x) \cos( j y) \cos( k z  ) + 
    a_{ijk}^{(2)} \cos( i x) \cos( j y) \sin( k z  ) + \nonumber \\
    &\qquad
    a_{ijk}^{(3)} \cos( i x) \sin( j y) \cos( k z  ) + 
    a_{ijk}^{(4)} \cos( i x) \sin( j y) \sin( k z  ) + \nonumber \\
    &\qquad
    a_{ijk}^{(5)} \sin( i x) \cos( j y) \cos( k z  ) + 
    a_{ijk}^{(6)} \sin( i x) \cos( j y) \sin( k z  ) + \nonumber \\
    &\qquad
    a_{ijk}^{(7)} \sin( i x) \sin( j y) \cos( k z  ) + 
    a_{ijk}^{(8)} \sin( i x) \sin( j y) \sin( k z  ) 
    \Big),
\end{align*}
where  $F(x, y, z)$ is periodic on $[0,2\pi]^3 \sim \mathbb{R}^3 / ( 2 \pi \mathbb{Z})^3$; each $a_{ijk}^{(p)}$, for $p=1,\dots,8$, is i.i.d.\ $\sim \mathcal{N}(0,1)$; and we take $M = 30$. In practice, to simulate voxellated data we evaluate $F$ at the centers of $N_v^3$ equal-size voxels by evaluating at discrete points $(x_l, y_m, z_n)=((2l-1)\pi/N_v,(2m-1)\pi/N_v,(2n-1)\pi/N_v)$ for integers $l,m,n=1,\ldots ,N_v$. The intensity of each grid point is given by the sampled value of $F$, rescaled linearly
to the 8-bit grayscale range $[0,255]$ (taking the nearest integer in the range).  
 
\begin{figure}[h!tbp]
    \centering
    \begin{subfigure}[b]{0.27\linewidth}
        \centering
        \includegraphics[width=\linewidth]{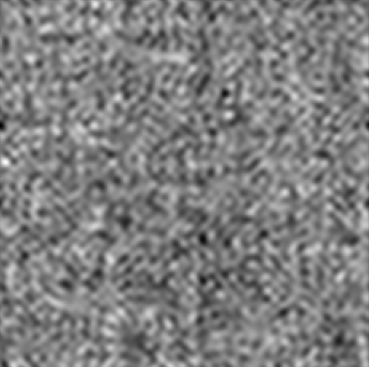}
        \caption{2D slice $N_v^2=254\times 254$ pixels}
        \label{fig:fourier_2d_slice}
    \end{subfigure}
    \hfill
    \begin{subfigure}[b]{0.3\linewidth}  
        \centering
        \includegraphics[width=\linewidth]{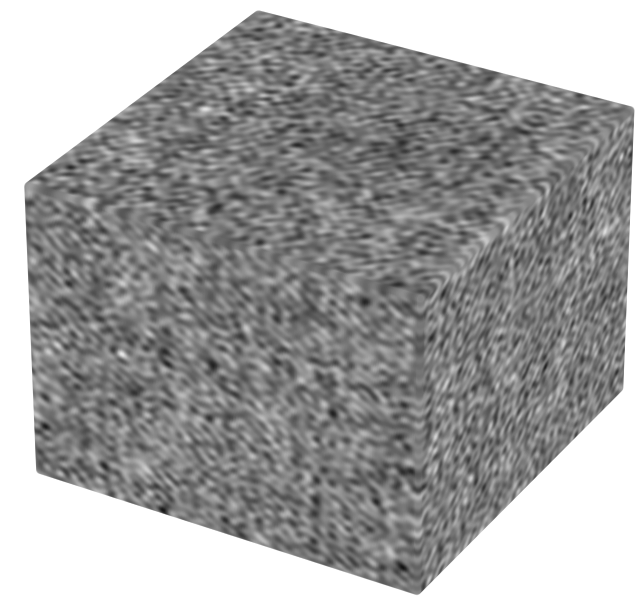}
        \caption{3D cube $N_v^3 = 254\times254\times254$ voxels}
        \label{fig:fourier_3d}
    \end{subfigure}
    \hfill
    \begin{subfigure}[b]{0.38\linewidth}
        \centering
        \includegraphics[width=\linewidth]{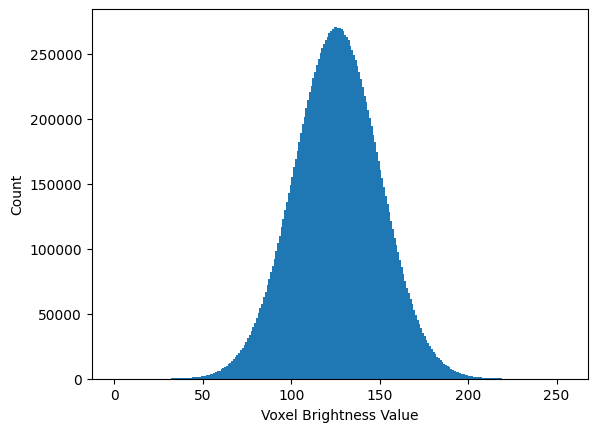}
        \caption{Distribution of voxel brightness values}
        \label{fig:fourier_values_distribution}
    \end{subfigure}
    
    \caption{Fourier dataset}
    \label{fig:fourier_comparison}
\end{figure}

 A 2D slice from the resulting dataset and the 3D representation can be seen in Figs.~\ref{fig:fourier_2d_slice} and \ref{fig:fourier_3d}. The random Fourier coefficients can produce a wide range of  patterns in the voxel intensities, and the resulting structure has features qualitatively similar to the pore-size variations and connectivity observed in 3D grayscale images of physical porous media. Fourier datasets are commonly studied as models for porous media in the materials literature~\cite{roding2020predicting}. 
 In terms of physical relevance, such datasets are perhaps most representative of crystalline porous materials; the periodic structures found in such media can be decomposed into the sum of principal wave modes, making them a prime candidate for Fourier analysis~\cite{ABID2024e23840}. Therefore, the Fourier dataset serves as a controlled test case for evaluating our method's ability to capture topological features across different scales. The grayscale histogram for the dataset described here follows a normal distribution, as seen in Fig.~\ref{fig:fourier_values_distribution}.

\subsection{PuMA Dataset}
\begin{figure}[h!tbp]
    \centering
    \begin{subfigure}[b]{0.27\linewidth}
        \centering
        \includegraphics[width=\linewidth]{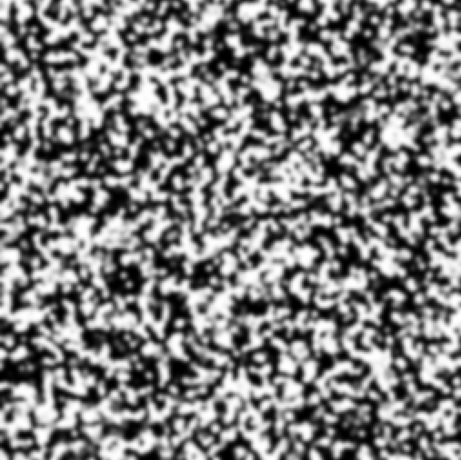}
        \caption{2D slice $n_v^2= 254\times 254$ pixels}
        \label{fig:puma_2d_slice}
    \end{subfigure}
    \hfill
    \begin{subfigure}[b]{0.3\linewidth}
        \centering
        \includegraphics[width=\linewidth]{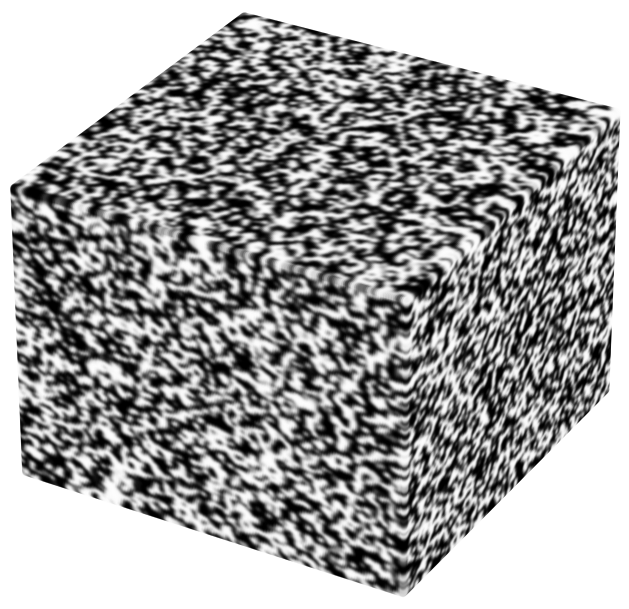}
        \caption{3D cube $N_v^3= 254\times 254\times 254$ voxels}
        \label{fig:puma_3d}
    \end{subfigure}
    \hfill
    \begin{subfigure}[b]{0.38\linewidth}
        \centering
        \includegraphics[width=\linewidth]{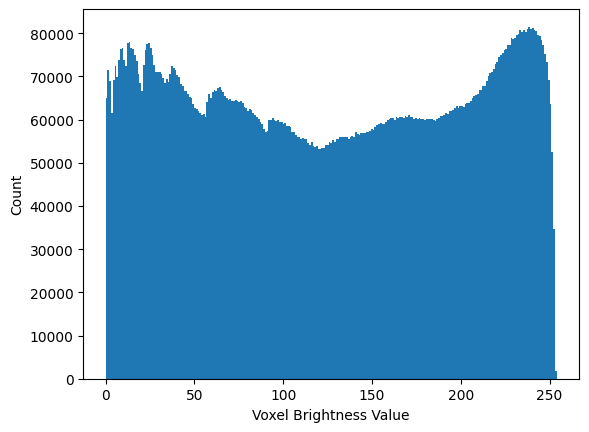}
        \caption{Distribution of voxel brightness values}
        \label{fig:puma_values_distribution}
    \end{subfigure}

    \caption{PuMA dataset}
    \label{fig:puma_dataset_visualizations}
\end{figure}

  The PuMA dataset was generated using the Porous Microstructure Analysis (PuMA) Library --- a free, open-source software developed by NASA~\cite{puma2021}. The PuMA library is widely-used, and was designed to import digital images of porous media obtained from $\mu$CT or to generate synthetic microstructures and quantify various morphological and physical properties of these structures. Our PuMA dataset is generated by randomly placing spheres within a cubic domain. All spheres were specified to have a diameter of $4$ voxels and were generated in a manner such that the dataset had an overall porosity of $0.5$ (with the overlapping spheres occupying 50\% of the total volume).  
  The simulated domain is discretized to create a $N_v^3$ voxel binary image in which voxels that are majority solid are assigned the value 255 and those that are majority void are assigned the value 0. Such binary datasets are, however, unrepresentative of the multi-modal distributions often found in experimental samples~\cite{https://doi.org/10.1029/2020GL088594}. To convert the image to grayscale, we perform a discrete convolution on the image with a Gaussian kernel of standard deviation $\sigma = 1.1$ as described in Sec.~\ref{Noising and Denoising}, obtaining a multi-modal grayscale distribution that emulates what one might find in experimental samples. A 2D cross-section and 3D visualization of the resulting PuMA dataset is shown in Figs.~\ref{fig:puma_2d_slice} and \ref{fig:puma_3d}, and the distribution of grayscale values in Fig.~\ref{fig:puma_values_distribution}. The final PuMA dataset may be considered analogous to many heterogeneous porous materials such as sandstone, ceramics, and other composite materials \cite{VANEEKELEN197375}. 

\subsection{Cellular Dataset}
  The  ``cellular'' dataset comes from the Porespy library --- an open-source Python toolkit designed for the quantitative analysis of images of porous materials \cite{Gostick2019PoreSpy}. The Porespy library provides python functions for generating artificial porous media, calculating geometric and topological metrics, and extracting pore networks. Underlying the Porespy library is the PyFastNoiseSIMD package, which implements high-performance noise generation, designed for procedural content generation in graphics and simulation applications \cite{pyfastnoisesimd}. We use this library to produce the Cellular dataset, which is based on Worley noise. 

  Worley noise is constructed from a set of feature points, or seeds. The package subdivides space into a lattice of congruent cubes --- the seed lattice --- and assigns exactly one seed to each cube. Each seed is displaced from the center of its lattice cube by a random offset. The maximum magnitude of this offset, measured as a fraction of the lattice cube side length, is controlled by the ``jitter'' parameter: a jitter of $0$ leaves every seed at the center of its cube, while larger values scatter the seeds more freely within their respective cubes. We employ a jitter value of $0.45$, which perturbs the seeds strongly without causing any wrapping artifacts during dataset generation. 

  From these seeds, we define a continuous scalar field over $\mathbb{R}^3$ in which each point is assigned the Euclidean distance to its nearest seed. This field partitions space into Voronoi cells --- one per seed, each consisting of the points closest to that seed --- whose mutual boundaries form the irregular, cell-shaped features of the dataset. To produce a discrete image, the field is sampled on the voxel grid of $N_v^3$ points. The voxel grid is distinct from the seed lattice; the former (finer) grid specifies where the scalar field is evaluated, while the latter governs where the Voronoi cells are located and how large they are. The relative scale of these two grids is set by a single parameter, termed ``frequency'' by the package: each lattice cube spans $1/\text{frequency}$ voxels along each axis, so an edge of $N_v$ voxels contains roughly $N_v \times \text{frequency}$ Voronoi cells. 
  A smaller frequency therefore makes each lattice cube larger relative to the voxel grid, so that every Voronoi cell spans more voxels and the resulting features appear coarser. The sampled distances in each voxel are then linearly mapped to grayscale intensity values within the range $[0, 255]$. For the Cellular dataset, we generate a volume of $N_v^3$ voxels with a frequency of $0.03$, chosen to produce relatively coarse cellular features.
 
\begin{figure}[h!tbp]
    \centering
    \begin{subfigure}[b]{0.27\linewidth}
        \centering
        \includegraphics[width=\linewidth]{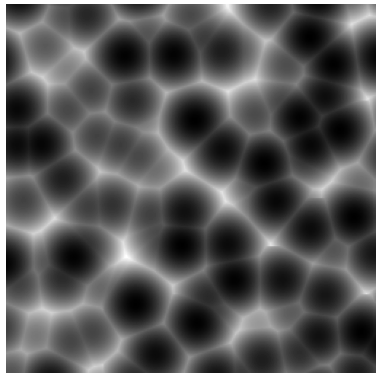}
        \caption{2D slice $N_v^2= 254\times 254$ pixels}
        \label{fig:cellular_2d}
    \end{subfigure}
    \hfill
    \begin{subfigure}[b]{0.3\linewidth}
        \centering
        \includegraphics[width=\linewidth]{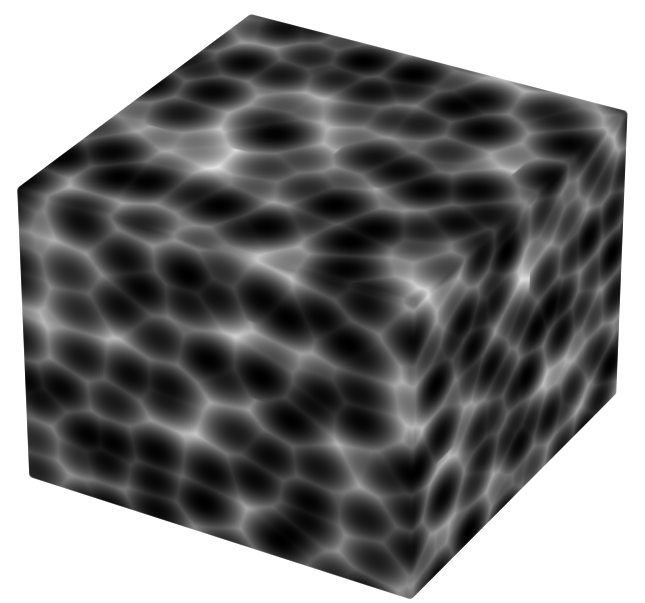}
        \caption{3D cube $N_v^3 = 254\times 254 \times 254$ voxels}
        \label{fig:cellular_3d}
    \end{subfigure}
    \hfill
    \begin{subfigure}[b]{0.38\linewidth}
        \centering
        \includegraphics[width=\linewidth]{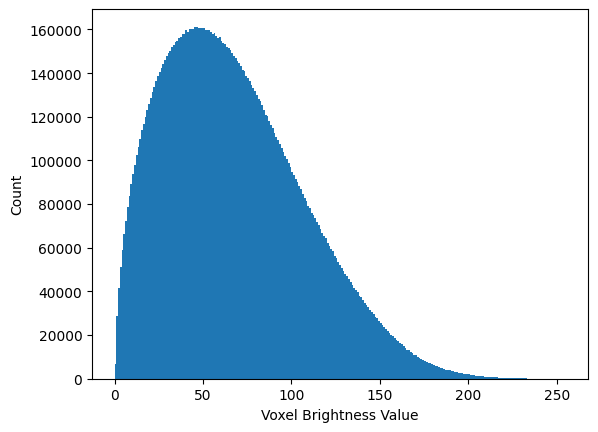}
        \caption{Distribution of voxel brightness values}
        \label{fig:cellular_values_distribution}
    \end{subfigure}

    \caption{Cellular dataset}
    \label{fig:cellular_comparison}
\end{figure}

  The Cellular dataset is topologically analogous to naturally occurring materials such as wood, cork, and bone, as well as man-made structures such as manufactured honeycombs and foams \cite{MACONACHIE2019108137}. The voxel brightness distribution for the Cellular dataset is skewed towards darker voxel values, allowing us to test our methodology on datasets that consist mostly of void space, which is common in $\mu$CT cross sections of fractured porous media samples \cite{niobrara, sandstone}. A 2D slice of the Cellular dataset is shown in Fig.~\ref{fig:cellular_2d}, a 3D visualization in Fig.~\ref{fig:cellular_3d}, and the distribution of grayscale voxel values in Fig.~\ref{fig:cellular_values_distribution}.

\section{Definitions of key topological concepts and measures} \label{Definitions}
  This section briefly introduces the key concepts from TDA that will be used throughout the rest of the paper. Section~\ref{sec:PH} below gives the key ideas in the context of this work, while Sec.~\ref{Persistence Tools} introduces the PH representations and topological measures that we will use to analyze the data. Readers interested in more detailed discussion of these concepts are referred to the book by Dey \& Wang~\cite{Dey_Wang_2022}. 

\subsection{Persistent Homology\label{sec:PH}} \label{Background}

The  homology groups of a space are classical topological invariants ~\cite{hatcher2002algebraic}. For our purposes, we need only concern ourselves with the homology groups arising from the geometric cubical complexes we encounter in three dimensional voxellated images \cite{mischaikow}, consisting of $0,1,2,3$ dimensional ``cubes'', corresponding to points, line segments, squares, cubes, respectively. 
For a given cubical complex $X$,  one may define a sequence of homology groups $ H_i(X )$ for natural numbers  $i$. Technically speaking, this definition depends on  the choice of scalar coefficients $\mathbb{F}$. For the purpose of defining persistent homology (PH), we take $\mathbb{F}$ to be a field;  for the results described in this paper, we take $\mathbb{F}$ to be $\mathbb{Z}/2\mathbb{Z}$. For each $i$, $ H_i(X) $ is isomorphic to a vector space $\mathbb{F}^{\beta_i}$, the dimension $\beta_i$ of which is referred to as the $i^{th}$ \emph{Betti} number. If $X \subseteq \R^n$, then $ \beta_i =0$ for all $ i \geq n$, hence for 3D data, we need only consider $ \beta_0, \beta_1, \beta_2$. 
Physically, these correspond to the number of connected components, loops, and cavities, respectively, in the imaged material, as discussed further below. 

Loosely speaking, PH describes how topology changes as one passes through a nested sequence of subsets. 
To fix ideas for the present context, consider a topological space  $ X $  (e.g. the voxellated unit cube with $N_v^3$ voxels) and a smooth function $ f: X \to \mathbb{R}$. 
Given such a function and a threshold $\theta$, we may define the super-level sets $ X_\theta = \{ x \in X : f(x) \geq \theta  \} $.  
For a given threshold, we may study the topology of $ {X}_\theta$, and compute  its homology groups; for every threshold $\theta$ we obtain a collection of  discrete algebraic invariants $(\beta_0,\beta_1,\beta_2)$: the number of connected components, loops, and cavities, respectively, that are present at this threshold.  

These superlevel sets naturally include into one another; if $ \theta_m < \theta_n $ then $ {X}_{\theta_m} \supseteq {X}_{\theta_n}$. 
In this manner  we may obtain  a {\it topological filtration} $\mathcal{F}$, i.e. a nested sequence of spaces $ {X}_{\theta_0} \supseteq {X}_{\theta_1} \supseteq \dots \supseteq {X}_{\theta_N}$, where $\theta_0<\theta_1< \ldots <\theta_N$. 
These nested subsets induce homomorphisms via inclusion mappings between their $i^{th}$ homology groups  $ H_{i} ({X}_{\theta_0} ) \leftarrow H_{i  } ({X}_{\theta_1} )  \leftarrow \dots \leftarrow H_{i } ({X}_{\theta_N} ) $, forming a persistence module. 
This may be decomposed into indecomposable intervals, identifying within which indexed space ${X}_b$ a homological generator first appears, and the first indexed space ${X}_d$ for which the generator is mapped to zero. Here $b$ is said to be this generator's \emph{birth} time, and $ d$ its \emph{death} time (here, `time' refers to the value of the filtration threshold $\theta$).

The $i$-dimensional PH of a filtration may thus be summarized as a collection of (not necessarily unique) birth-death intervals $\{ I_i^j \}_{j \in J} =\{ [b^j_i,d^j_i]\}_{j \in J} $ for some index set $J$. 
More precisely, if a specific homology generator first appears in $H_i(X_b)$, and is first mapped to zero by the inclusion induced map  $\iota_i:H_i(X_b) \to H_i(X_d)$, then $[b,d]$ is referred to as the corresponding $i$-dimensional \emph{persistence interval}. 
The $0$-dimensional PH tracks when connected components first appear and later merge together; 
the $1$-dimensional PH tracks when loops in the space appear and disappear; and the 2-dimensional PH tracks cavities. 

\subsection{Representations of Persistent Homology and Topological Measures} \label{Persistence Tools}

As defined above, the $i$-dimensional PH of a 3D image is completely described by an (unstructured)  collection of intervals. 
In this work we use several tools that allow us to visualize, summarize, and quantify the topological information gathered on our datasets, loosely referred to as topological measures, which we define formally and informally below. 
 
\begin{figure}[t!]
    \centering
    \begin{subfigure}[b]{0.32\linewidth}
        \centering
        \includegraphics[width=\linewidth]{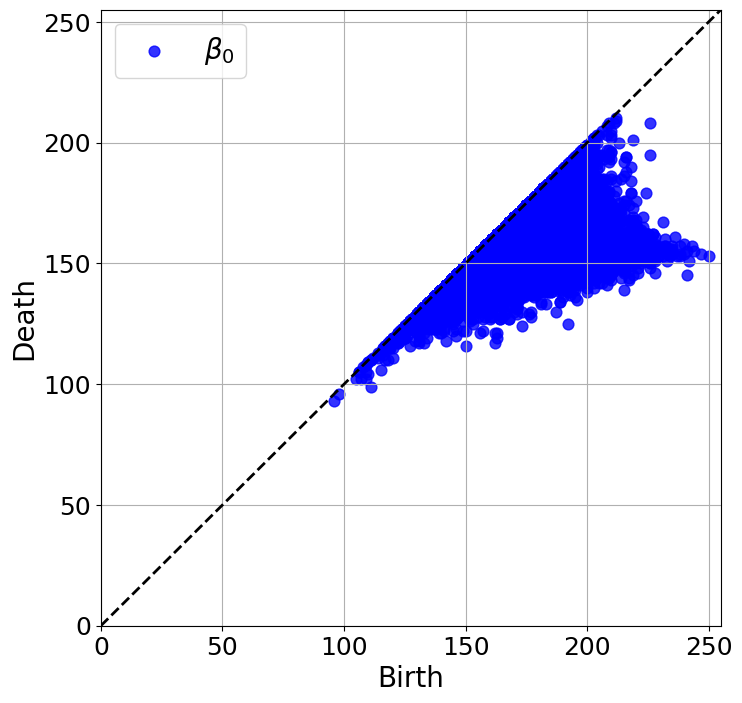}
        \caption{Persistence diagram}
        \label{fig:pd_example}
    \end{subfigure}
    \hfill
    \begin{subfigure}[b]{0.32\linewidth}
        \centering
        \includegraphics[width=\linewidth]{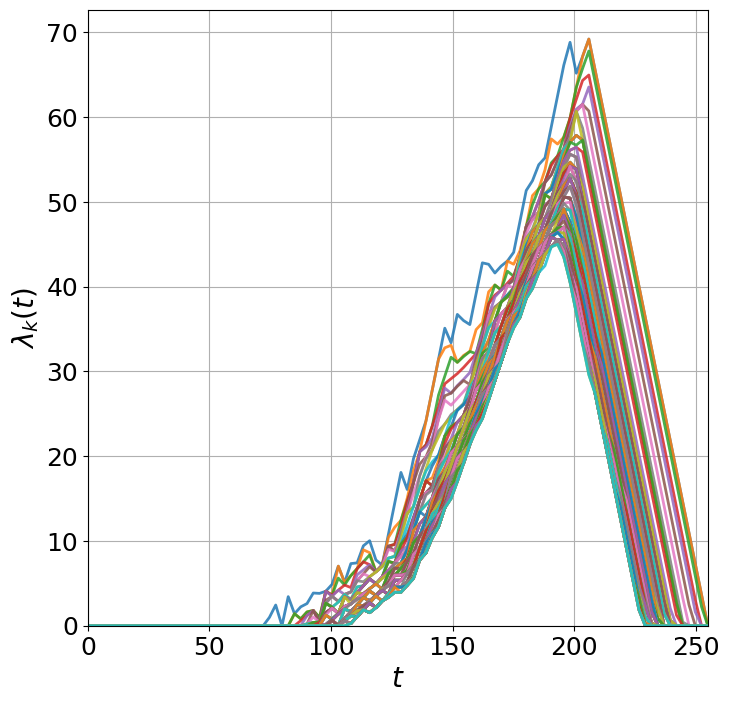}
        \caption{{Persistence landscape}}
        \label{fig:pl_example}
    \end{subfigure}
    \hfill
    \begin{subfigure}[b]{0.32\linewidth}
        \centering
        \includegraphics[width=\linewidth]{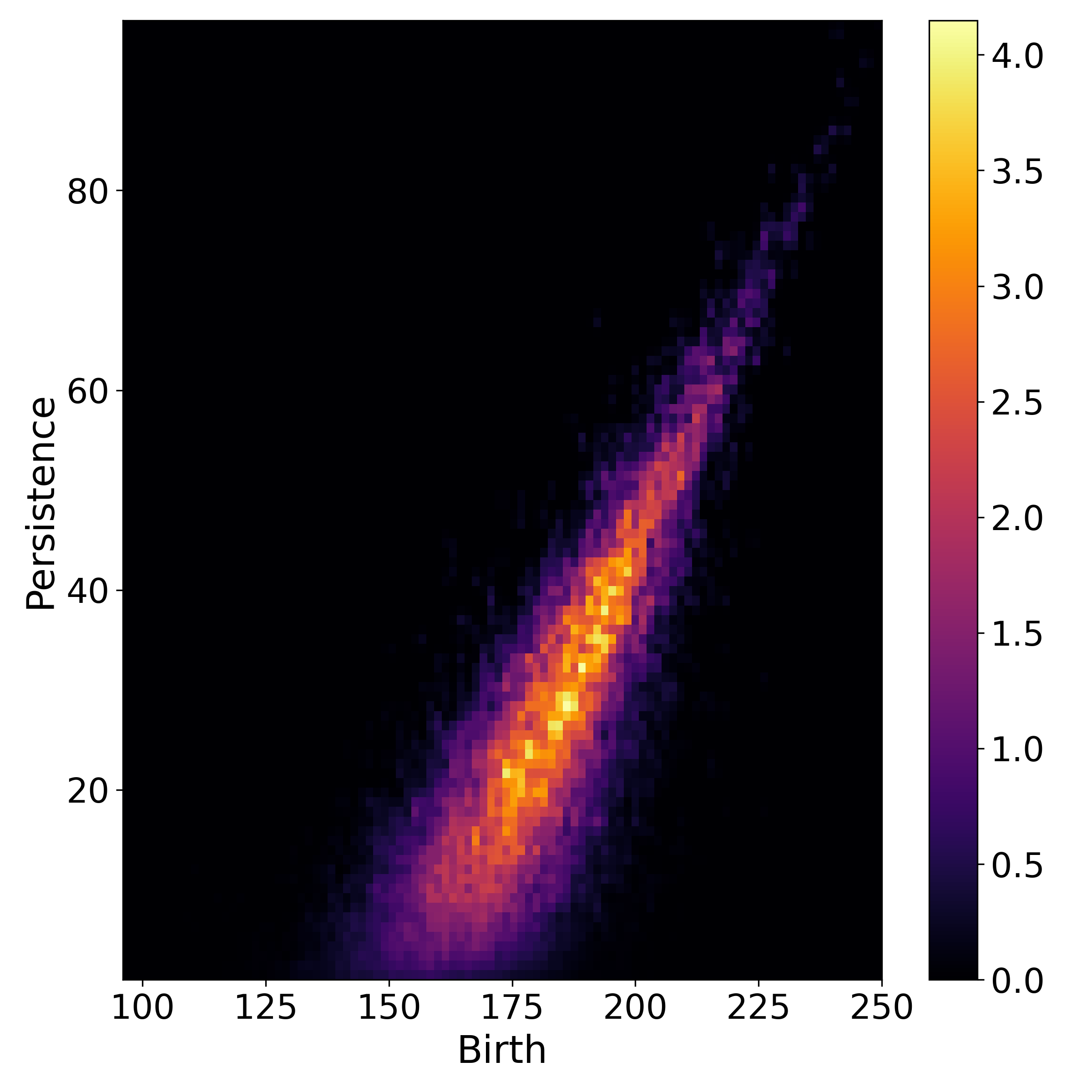}
        \caption{{Persistence image }}
        \label{fig:pi_example}
    \end{subfigure}
        \caption{Three representations of the 0-dimension ($\beta_0$) topological information for the Fourier dataset. Fig.~\ref{fig:pd_example} shows the birth and death values of connected components as the grayscale filtration parameter varies (the PD). Each point corresponds to a generator, with horizontal distance to the diagonal representing its  
        lifespan (sometimes also called the persistence of the generator). Fig.~\ref{fig:pl_example} represents the data as a collection of functions created from the birth and death values of generators (the persistence landscape; see Definition~\ref{def:Persistence Landscape}). Fig.~\ref{fig:pi_example} shows a smoothed and discretized representation of the persistence diagram (the persistence image). The color value of each pixel corresponds to the integral of weighted Gaussian kernels centered at the birth–persistence locations of generators; see Definition~\ref{def:persistence-image}.
               }
\label{fig:persistent_graph_examples}
    
\end{figure}

A primary tool used to understand the topological features of a given 3D image is the Persistence Diagram, $\rm{PD}(\mathcal{F})$, a multi-subset of $ \R^2$ that depends on the chosen filtration $\mathcal{F}$. In essence, a PD contains the birth-death pairs $(b_i,d_i)$ of generators that belong in $H_i$ where $i=0,1,2$ as we move through $\mathcal{F}$ and encounter the birth and death of each specific generator; see Fig.~\ref{fig:pd_example} for an example representing the 0-dimensional topology ($i=0$) of the Fourier dataset. The diagonal of a PD represents trivial features (zero lifespan); the distance of a point from the diagonal represents its topological persistence, and points near the diagonal are often considered to be ``noise''  \cite{Edelsbrunner_Letscher_Zomorodian_2002}. 

The persistence barcode is another method for visualizing PH, and is a particularly useful way to visualize the impact that noise has on a dataset.   The idea behind the persistence barcode is to represent each birth-death interval $(b_j,d_j)$ as a stacked collection of horizontal bars (like a barcode), where each bar's horizontal extent ranges from $b_j$ to $d_j$ \cite{Ghrist2008}. In our work we produce persistence barcodes for $i=0,1,2$, see Fig. \ref{fig:fourier_barcode_all} later for a specific example.

Using these concepts, we now define several topological measures on our original and denoised datasets. 
The primary motivation is to assess whether each measure is able to retain the essential topological features of the original 3D image after the noising-denoising process. 
With this in mind, we create normalized measures that provide an objective indication of how different the original and denoised datasets are. Important features to consider include the number of generators present within the datasets and the average lifespans of those generators (with respect to our chosen grayscale-level filtration).

The first and perhaps simplest measure is based on the number of topological generators in the original and denoised datasets, which are just the total number of generators that are encountered as we move through a filtration $\mathcal{F}$. Let $N_i^o$, $N_i^d$ be the number of generators belonging to the $i$-th homology group for the true (original) dataset and the denoised dataset. The normalized difference between these numbers, 
    \begin{equation}
        \Delta N_i = |N_i^o-N_i^d|/N_i^o,
        \label{Number of Generators}
    \end{equation}
    provides one simple measure of how different the datasets are.
We refer to this difference as the \emph{number of generators} measure.

A second simple measure that we introduce pertains to the average lifespan of generators. Let $(b_i,d_i)$ be a birth-death pair for a single generator belonging to the $i$-th homology group in a PD. The lifespan $L_i$ of the generator is $L_i = |d_i-b_i|$, and the average lifespan of the $N_i$ generators within the PD, $\bar{L}_i$, is
    \begin{equation*}
        \bar{L}_{i}= \frac{1}{N_i}\sum_{j=1}^{N_i}{L_i^j}.
    \end{equation*}
For the purpose of this work, we are interested in the absolute (normalized) difference between the average lifespans of generators in the original dataset, $\bar{L}^o_i$, and in the denoised dataset, $\bar{L}_i^d$, i.e.
\begin{equation}
        \Delta L_i = |\bar{L}_{i}^o-\bar{L}_{i}^d|/\bar{L}_i^o.
        \label{Average Lifespan of Generators Measure}
\end{equation}
We refer to this difference as the \emph{average lifespan} measure.

We also use several standard distances in TDA to create a set of distance measures between the PDs of our original and denoised datasets. These include the bottleneck and Wasserstein distances, each of which provide a measure of the similarity between two PDs.  
\begin{definition} [Bottleneck Distance] \label{Bottleneck Distance}
Consider two persistence diagrams ${\rm PD}_i^1$ and ${\rm PD}_i^2$ corresponding to the $i$-th homology group. 
The bottleneck distance between the two is defined as
\begin{equation}
\label{eq:Bottleneck_Distance}
d_b({\rm PD}_i^1, {\rm PD}_i^2) := \inf_{\pi:\rm{PD}_i^1\to\rm{PD}_i^2}{\sup_{x \in \rm{PD}_i^1}{|x-\pi(x)|}},
    \end{equation} 
where $\pi$ ranges over all bijections between points in ${\rm PD}_i^1$ and ${\rm PD}_i^2$, allowing points to be matched to the diagonal if necessary.
\end{definition} 
A key utility in using this  distance metric comes from  the bottleneck stability theorem \cite{StablityPersistenceDiagrams}, 
which guarantees that if the distance between two functions is small in the supremum norm, then the bottleneck distance between their corresponding  PDs (induced by the sub/superlevelset filtrations of the functions) is also small. Put more precisely, and in the context of this work, if we consider two persistence diagrams PD$_i^1 (f_1)$ and PD$_i^2 (f_2)$ arising from filtrations $f_1$ and $f_2$ on the cubical complexes $X_1$ and $X_2$,  (in our application $X_1$, $X_2$ will be the original and denoised datasets, and $f_1, f_2$ the grayscale voxel functions for each of these sets; see Sec.~\ref{sec:PH}),   then  
\begin{equation}
    d_b({\rm PD}_i^1(f_1), {\rm PD}_i^2(f_2)) \le \|f_1-f_2\|_{\infty},\label{eq:BottleneckStabilityThm}
\end{equation}
where $\|\cdot\|_\infty$ denotes the supremum norm.

If we view the bottleneck distance as providing a sort of $L^\infty$ distance on the space of persistence diagrams, then the $L^p$  analogue is the $p$-Wasserstein distance  
\cite{Dey_Wang_2022}: 
\begin{definition} [Wasserstein Distance] \label{Wasserstein Distance}
Consider two persistence diagrams ${\rm PD}_i^1$ and ${\rm PD}_i^2$ corresponding to the $i$-th homology group. The $p$-Wasserstein distance is defined as
\begin{equation*}
    W_p({\rm PD}_i^1, {\rm PD}_i^2) = \left( \inf_{\pi: {\rm PD}_i^1 \to {\rm PD}_i^2} \sum_{x \in {\rm PD}_i^1} \|x - \pi(x)\|_\infty^p \right)^{1/p},
\end{equation*}
where $\pi$ ranges over all bijections between points in ${\rm PD}_i^1$ and ${\rm PD}_i^2$, allowing points to be matched to the diagonal if necessary. Here, $p \ge 1$ determines the order of the Wasserstein distance.
\end{definition}

The Wasserstein distance also enjoys a stability  \cite{skraba2025wasserstein,StablityPersistenceDiagrams,cohen2010lipschitz}, however, as Skraba and Turner note, it ``\emph{appears to be one of the most misunderstood and miscited results within the field of topological data analysis. Common errors include using a small p (often 1 or 2) for high dimensional data [and] assertions that the persistence diagrams depend Lipschitz-continuously on data}''~\cite{skraba2025wasserstein}. 
In this work we only study ${\rm PD}_i$ for $ i=0,1,2$ and employ $p=2$ for the Wasserstein order; investigating results for other values $p$  would be interesting, but is beyond the scope of this paper. To summarize, we are specifically interested in calculating $d_b({\rm PD}_i^o, {\rm PD}_i^d)$, the bottleneck distance between the original and denoised PDs of our datasets for each homology group $H_i$, and $W_2 ({\rm PD}_i^o,{\rm PD}_i^d)$.  

From a PD we can generate other representations of the topological composition of 3D images such as persistence landscapes \cite{bubenik2020} and persistence images \cite{adams2017}. These constructions are useful for statistical analysis since they map PDs to a vector space, allowing for the creation of vectorized persistence summaries that can be directly input into machine learning prediction pipelines \cite{ali2023survey}. 

\begin{definition} [Persistence Landscape] \label{def:Persistence Landscape}
Given a finite persistence diagram ${\rm PD}_i = [(b_i^j, d_i^j)]_{j\in[1,N_i]}$ containing $N_i$ birth-death pairs corresponding to the $i$-th homology group, the persistence landscape with respect to ${\rm PD}_i$ is the function $\lambda_{{\rm PD}_i}: \mathbb{N} \times \R \rightarrow \R$ where
\begin{equation} \label{Landscape}
\lambda_{{\rm PD}_i}(t) := \max\{[\min\{t-b_i^j, d_i^j-t\}], 0\} \text{ for } j \in [1,N_i].
\end{equation}
Moreover, we define the $k$-th persistence landscape, $\lambda_{{\rm PD}_i}^k$, to be the $k$-th largest value of \eqref{Landscape}. Specifically, for fixed $k$, $\lambda_{{\rm PD}_i}(t): \R \rightarrow \R$ where each persistence point $(b_i^j, d_i^j)$ gives rise to an upward pointing triangle defined as 
\begin{equation*}
[(t, \max\{\min\{t-b_i^j, d_i^j-t\}, 0\}) \space | \space t \in \R],
\end{equation*}
where the $\lambda_{\rm PD}^k$ is the $k$-th upper envelope in the arrangement formed by the union of these triangles \cite{Dey_Wang_2022}.
\end{definition}

  The persistence landscape is the transformation of a PD's persistence data into a functional form by converting birth-death pairs into a sequence of piecewise-linear functions that, when combined, form the persistence landscape~\cite{bubenik2020}. We can simply discretize these linear functions to create a vectorized representation of the persistence landscapes. A specific example of a persistence landscape is shown in Fig.~\ref{fig:pl_example} for $1\leq k \leq 100$.

Another vectorization of a PD is a persistence image, which is obtained by placing a Gaussian kernel at each point in the PD and combining them to form a smooth surface~\cite{adams2017}; a specific example is shown in Fig.~\ref{fig:pi_example}. More precisely, we define:
\begin{definition}[Persistence Surface \cite{Dey_Wang_2022,   adams2017}]
\label{def:persistence-image}
Let ${\rm PD}_i$ be a  
persistence diagram with finitely many points for the $i$-th homology group, and let
$T:\mathbb{R}^2\to\mathbb{R}^2$ be the linear transformation
\[
T(x,y)=(x,x-y),
\]
which maps birth--death coordinates to birth--lifespan coordinates. Let
$\phi:\mathbb{R}^2\to\mathbb{R}$ be a differentiable probability distribution and let
$\omega:\mathbb{R}^2\to\mathbb{R}$ be a non-negative weight function. Let
$T({\rm PD}_i)$ denote the transformed persistence diagram. The persistence surface
$\mu_{{\rm PD}_i}:\mathbb{R}^2\to\mathbb{R}$ is defined by
\[
\mu_{{\rm PD}_i}(x,y)
=
\sum_{z\in T({\rm PD}_i)}
\omega(z)\phi_z(x,y).
\]
\end{definition}
Typically $\phi$ is taken as a  normalized Gaussian distribution $N(\mu, \sigma^2)$, defined as 
\begin{equation*}
\phi(x, y) = \frac{1}{2\pi\sigma^2} e^{-\frac{(x^2+y^2)}{2\sigma^2}} \text{ for } (x,y) \in \R^2,
\end{equation*}
while for a given threshold $\theta$ and power $m$,  a common choice of  weight function is 
\begin{align} \label{eq:WeightFunction}
\omega=  \omega_{\theta,m} (x,y) &=
\begin{cases}
 0  & \mbox{if } |x-y| = 0 \\
 \big(\frac{|x-y|}{\theta}\big)^m  & \mbox{if } 0<|x-y| < \theta \\
 1  & \mbox{if } |x-y| \geq \theta .
\end{cases}  
\end{align}
This weight function serves to discount points whose lifespan is less than $\theta$. This smooth surface $\mu_{{\rm PD}_i}$  can be discretized to create an image that is a vectorized representation of a PD. 

In our implementation, persistence surfaces are represented in birth--lifespan coordinates (recall that, for the superlevel-set filtrations considered in this work, 
lifespan $L$ is computed as $L=|d-b|$, where $b$ and $d$ denote the birth and death values of a feature, respectively). The persistence surface is constructed using a normalized Gaussian kernel with variance parameter $\sigma=0.5$. We employ the piecewise linear weight function $\omega_{\theta,m}$ defined in (\ref{eq:WeightFunction}) with $m=1$ and $\theta=L_{\max}$, where $L_{\max}$ denotes the maximum lifetime value in the PD. Consequently, features are weighted linearly according to their lifetimes, assigning greater importance to longer-lived topological features.

\begin{definition}
    The persistence image, $I_{{\rm PD}_i}$, of the persistence diagram ${\rm PD}_i$ is the discretization of this persistence surface, where we fix $N, M > 0$ to create a collection $K$ of $M\times N$ rectangles that create a gridspace for ${\rm PD}_i$ and define $I_{{\rm PD}_i}$ as the following collection of integrals over all rectangles $\kappa\in K$: 
\begin{equation*}
I_{{\rm PD}_i} = \{I_{{\rm PD}_i}[\kappa]\}_{\kappa \in K}, \quad \text{where} \quad  I_{{\rm PD}_i}[\kappa] := \int_{\kappa} \mu_{{\rm PD}_i}(x,y)\, dydx.
\end{equation*}
\end{definition}

\section{Methods and Preliminary Results}
\label{sec:methodology}

In this section, we describe the pipeline schematized in Figure \ref{fig:Testing Process}. First, in Sec.~\ref{Noising and Denoising}, we outline how we add noise to our original datasets to produce noisy datasets, and discuss the Gaussian smoothing method used to denoise them. 
  We then present and describe some preliminary results that highlight features of our methodology and provide context for our main results, which are described in Sec.~\ref{Results}. 

\subsection{Noising and Denoising Process} \label{Noising and Denoising}

  After generating each of our datasets as described in Sec.~\ref{Datasets}, we create a noisy version of each by introducing spatially independent noise to our original datasets. 
 The noise added to each voxel is sampled i.i.d. from a normal distribution $N(\mu,\sigma)$, where $\mu$ is the mean and $\sigma^2$ is the variance.  In our analysis we take $\mathbf{\mu}=0$ and $\sigma=5.76$.\footnote{This value of $\sigma=5.76$, expressed in units of voxel brightness, was obtained by analyzing experimental data ($\mu$CT images of coquina rock, provided by Marcio S. Carvalho of PUC Rio, Brazil) to determine the distribution of voxel brightness values from ``blank'' areas of the image.} We sample from this probability density function $N_v^3$ times independently for each voxel center $(x,y,z)$ to create a noise array of shape $N_v \times N_v \times N_v$, which we add to the original dataset voxel-wise to produce our noisy dataset. To ensure that brightness values remain within the range $[0,255]$, we truncate values below $0$ and above 255 back to $0$ and $255$, respectively. 

After producing our $N_v^3$ voxel noisy image, we denoise it using a Gaussian filter; we convolve the noisy image  with  a Gaussian:
\begin{equation} \label{Gaussian}
N(\mathbf{r}; \mathbf{\mu}, \sigma^2) = \frac{1}{\sqrt{(2\pi\sigma^2)^3}}\exp\left(-\frac{\|\mathbf{r}-\boldsymbol{\mu}\|_2^2}{2\sigma^2}\right), \qquad \mathbf{r}=(x,y,z),
\end{equation}
where we take $\boldsymbol{\mu} = (0, 0, 0)$ and vary our denoising level $\sigma$ on the interval $[0,10]$. We renormalize the resulting values post-convolution to integers on the interval $[0, 255]$ by (i) truncating floating point values to integers after each step of the discrete convolution and (ii) wrapping around values outside of $[0, 255]$ back to the interval in a periodic fashion i.e. values above $255$ will overflow back to $0$ and values below $0$ will underflow back to $255$. Applying this process to our noisy images results in an $N_v^3$ voxel denoised image. 

While naive Gaussian smoothing provides a computationally inexpensive baseline for noise reduction, its symmetric nature inherently blurs high-frequency structural boundaries, leading to poor edge preservation. Furthermore, it requires empirical tuning of the parameter $\sigma$ to adapt to different noise profiles. To mitigate these limitations, we also explore an alternative machine learning (ML) based approach, employing a U-Net architecture designed to learn a robust, non-linear mapping from the noisy inputs to reconstruct the original, noise-free images. 

The U-Net \cite{Ronneberger2015UNet} is a fully convolutional neural network that has become a standard architecture for image-to-image learning tasks. It consists of two symmetric paths. The first is an encoder: a sequence of convolutional layers interleaved with pooling operations that progressively down-sample the image's spatial resolution, yielding a coarse latent representation of its essential features. The second is a decoder, which progressively up-samples this latent representation back to the original resolution. Because the encoder's downsampling discards fine positional detail, such as the precise location of edges, the decoder cannot recover these features on its own. The U-Net addresses this through skip connections: at each resolution level, the feature maps produced by the encoder are passed directly to the corresponding decoder level and concatenated with the upsampled features there. These connections reintroduce the precise voxel-level positions of edges and local features that would otherwise be lost during downsampling, allowing the network to combine fine-grained local geometry with the global context captured at coarser scales. We employ a 3D U-Net \cite{Cicek20163DUNet} for our image volumes, which replaces the two-dimensional operations of the original architecture with their three-dimensional counterparts.

The ML model was optimized using a Huber loss function \cite{10.1214/aoms/1177703732, terven2025comprehensive}, which balances the Mean Squared Error (MSE) and Mean Absolute Error (MAE) between the original and denoised images. For small errors, the loss operates as MSE---the Maximum Likelihood Estimator in the presence of Gaussian noise---which ensures stable convergence. For large errors, it transitions to an MAE loss, which limits the influence of outliers and fundamentally aids in preserving sharp structural edges. For each of the three morphological datasets, the model was trained and evaluated on a set of 450 image volumes of dimension $N_v^3 = 128 \times 128 \times 128$ -- the smaller number of voxels in the images was chosen due to computational constraints and to enable a larger sample size for training. Each dataset was partitioned using a standard 64\%--16\%--20\% (train--validation--test) split, yielding 288 volumes for training, 72 for validation to monitor overfitting, and 90 for the final hold-out testing phase.

\subsection{Preliminary Results for Gaussian Denoising\label{preliminary}}

We now present some preliminary results, focusing on Gaussian denoising of the Fourier dataset, to illustrate key concepts and help contextualize our main results that follow in Sec.~\ref{Results}. Persistent homology computations were performed using the cubical complex implementation provided by the Cubicle software package \cite{wagner2023slice}. 

\begin{figure}[h!tbp]
    \centering
    \includegraphics[width=0.9\linewidth]
    {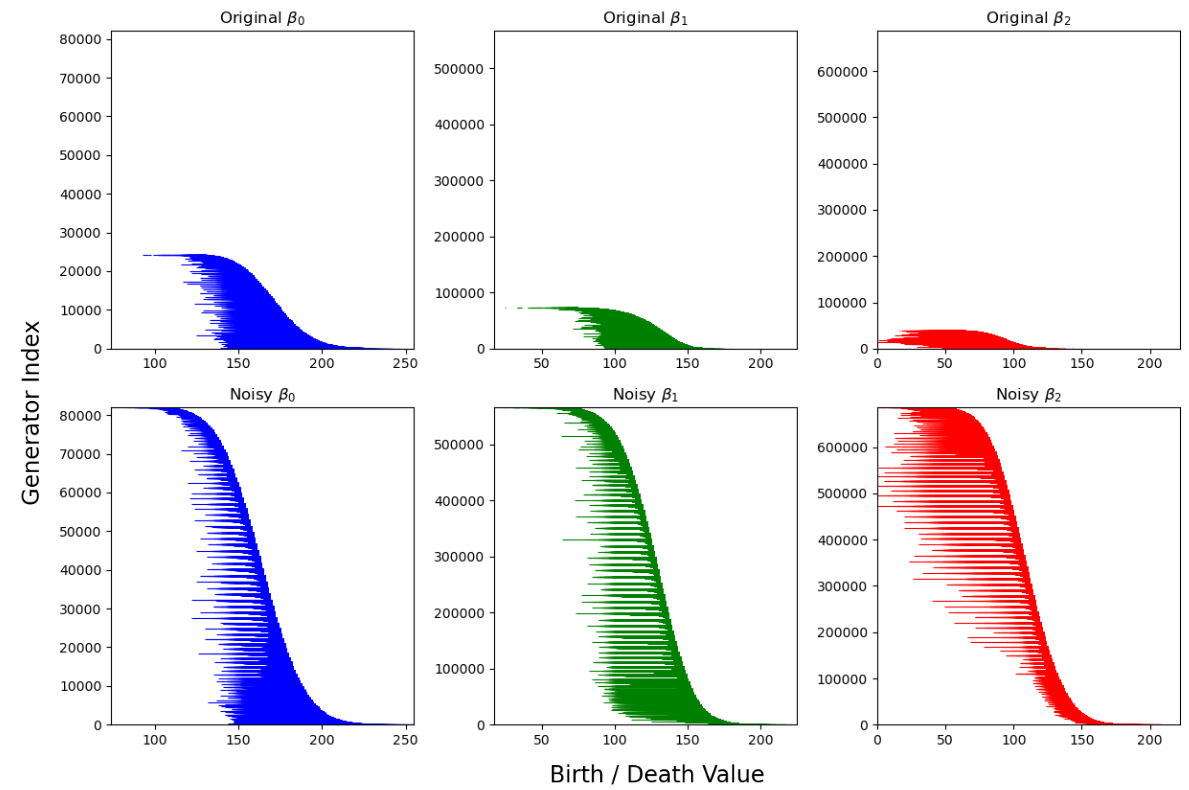}
    \label{fig:fourier_barcode}
    \caption{Persistence barcodes for the dimension 0, 1 and 2 PH ($\beta_0$, $\beta_1$, and $\beta_2$) for the original and noisy Fourier datasets. The horizontal axis represents the birth/death grayscale value. With the superlevel thresholding that we use, the right end of a line represents the grayscale value at which a feature is born, and the left end its death value. The vertical axis is an index number for each feature; here features are ordered by birth. The maximum vertical coordinate value corresponds to the total number of features, or topological generators.}
    \label{fig:fourier_barcode_all}
\end{figure}

We first highlight the effect of adding noise to our raw datasets. Introducing experimental noise to an image leads to a sharp increase in the number of (topologically) short-lived features, which may be visualized through persistence barcodes. As described in Sec.~\ref{Persistence Tools}, persistence barcodes represent each feature within a given homology group as a bar whose length corresponds to the feature's lifespan under filtration: long bars represent more persistent and essential features, while short bars represent short-lived features. In Fig.~\ref{fig:fourier_barcode_all}, we compare the persistence barcodes for the original and noisy Fourier datasets for $\beta_0,\beta_1,$ and $\beta_2$ generators. We see that the noise introduces a large number of short bars in the persistence barcodes, and thus a much larger vertical spread for the noisy dataset than for the original one. Table~\ref{Table: Number of Generators} shows specifically how the number of generators of each type increases dramatically on the addition of noise (for completeness, results for all three datasets are shown).

\begin{table}[h!tbp]
\begin{center}
\begin{tabular}{||c r r r||} 
    \hline
    Original Dataset & $N_0$ & $N_1$ & $N_2$ \\ [0.5ex] 
    \hline\hline
    Fourier & 24,298 & 72,777 & 39,567 \\ 
    \hline
    PuMA & 32,671 & 150,417 & 84,390 \\
    \hline
    Cellular & 3,159 & 11,985 & 402 \\
    \hline
\end{tabular}
\hfill
\begin{tabular}{||c r r r||} 
    \hline
    Noisy Dataset & $N_0$ & $N_1$ & $N_2$ \\ [0.5ex] 
    \hline\hline
    Fourier & 82,049 & 566,896 & 686,605 \\ 
    \hline
    PuMA & 73,354 & 330,608 & 338,301 \\
    \hline
    Cellular & 168,941 & 1,321,211 & 1,395,067 \\
    \hline
\end{tabular}
\caption{Total number of generators of each type in our three datasets prior to and after the addition of noise. } \label{Table: Number of Generators}
\end{center}
\end{table}
 
 We preface our main results by discussing the general trend that a typical denoising method should produce in our measures. We anticipate a relatively large error between the original and noisy dataset. As we begin to denoise the noisy dataset, increasing the value of the denoising parameter $\sigma$, the  error should decrease to some minimum value as we approach some optimal value of $\sigma$, before increasing again once we pass this value. Since all of our measures are normalized by the measure for the original dataset, we expect the normalized error between the denoised and original dataset to asymptote to 1 as $\sigma \to \infty$ and the denoised image becomes featureless.
 
\begin{figure}[h!tbp]
    \centering
    \begin{subfigure}[b]{0.32\linewidth}
        \centering
        \includegraphics[width=\linewidth]{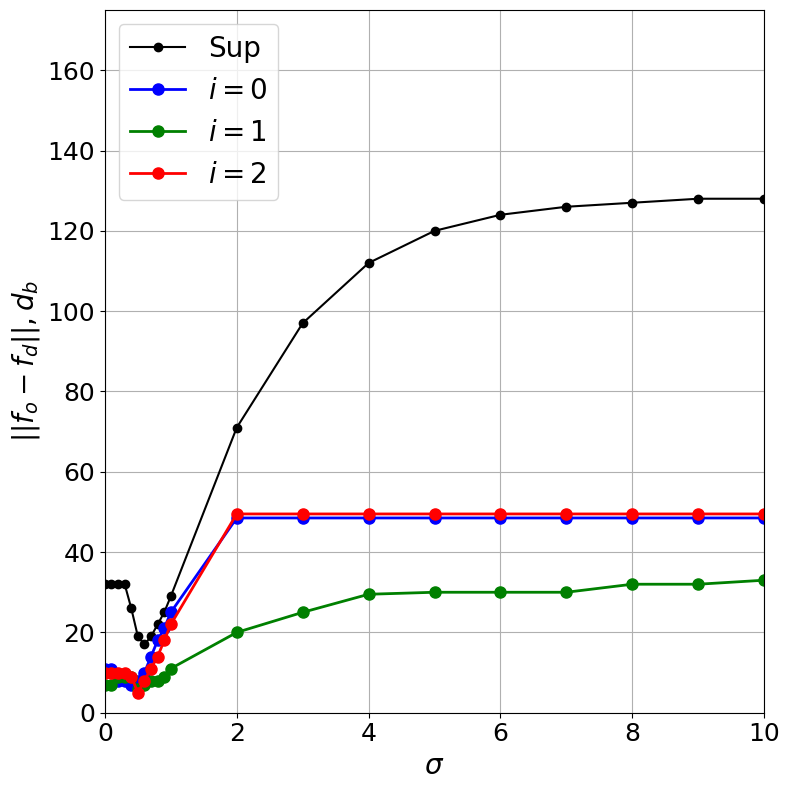}
        \caption{Fourier}
        \label{fig:fourier_supnorm_dist}
    \end{subfigure}
    \hfill
    \begin{subfigure}[b]{0.32\linewidth}
        \centering
        \includegraphics[width=\linewidth]{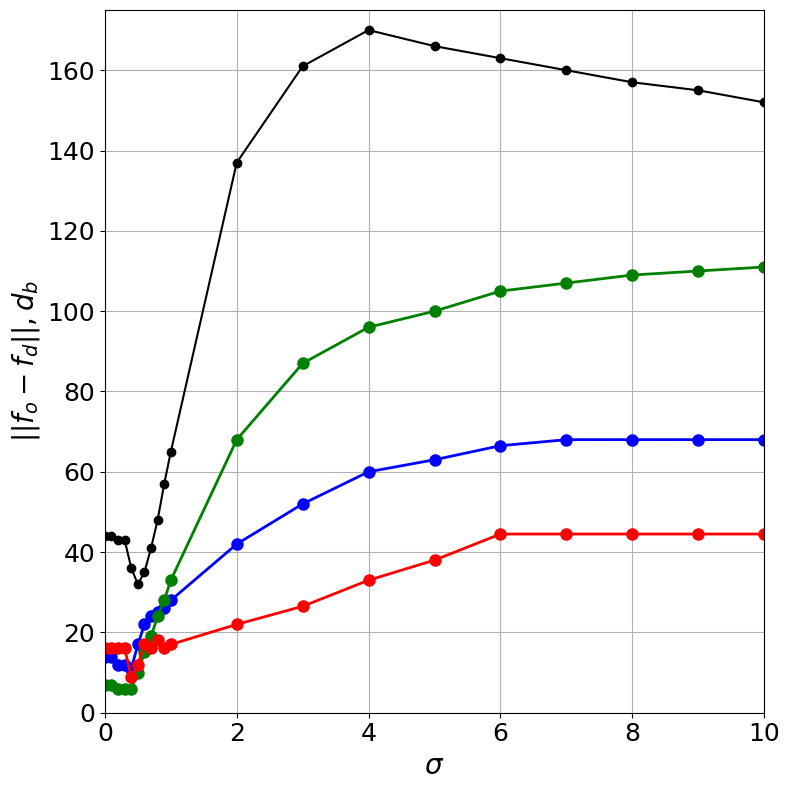}
        \caption{PuMA}
        \label{fig:puma_supnorm_dist}
    \end{subfigure}
    \hfill   
    \begin{subfigure}[b]{0.32\linewidth}
        \centering
        \includegraphics[width=\linewidth]{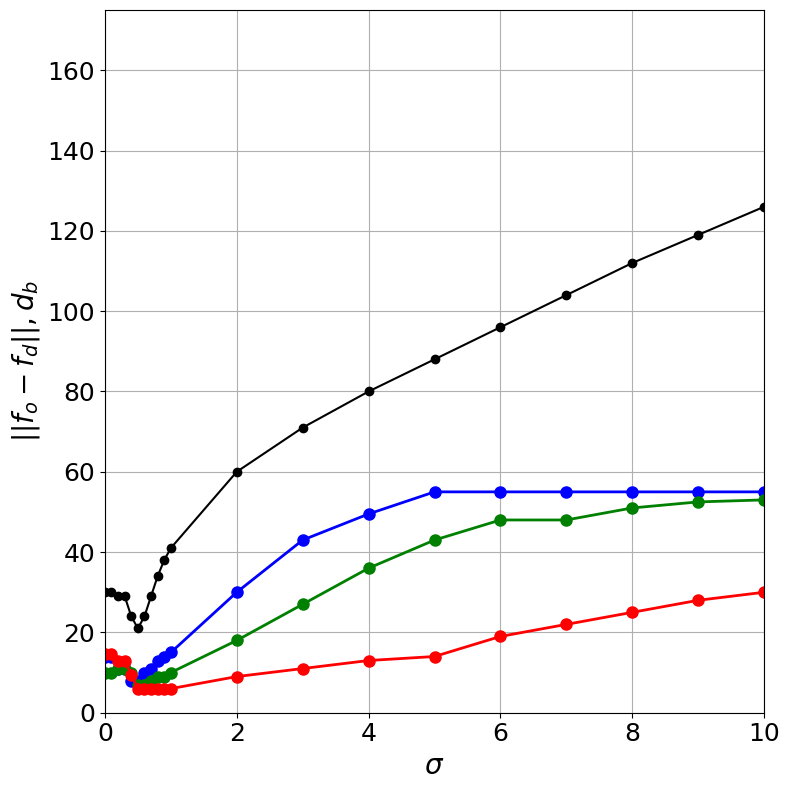}
        \caption{Cellular}
        \label{fig:cellular_supnorm_dist}
    \end{subfigure}

    \caption{ 
        Visualization of the Stability Theorem of Persistent Homology. Shown in black is $\|f_o-f_d\|_{\infty}$, the supremum norm distance between the grayscale thresholding functions used to induce the filtrations on the original and denoised datasets, while the non-normalized bottleneck distance measures $d_b ({\rm PD}_i^o, {\rm PD}_i^d)$ between the PDs for the original and denoised datasets for the dimension $i=0,1,2$ PH are shown in blue, green, and red respectively (see Definition \ref{Bottleneck Distance}). }
    
    \label{fig:supnorm_all}
\end{figure}

 To demonstrate the denoising process, we present our methodology in the context of  the bottleneck stability theorem ~(cf. Eq. \eqref{eq:BottleneckStabilityThm}). Specifically, we check that the non-normalized bottleneck distance $d_b$, defined in Eq.~\eqref{eq:Bottleneck_Distance}, between the denoised and original PDs of a dataset, is bounded by the supremum norm measure of the difference between the grayscale brightness functions for these datasets \cite{StablityPersistenceDiagrams}.    
  Figs.~\ref{fig:fourier_supnorm_dist},~\ref{fig:puma_supnorm_dist} and \ref{fig:cellular_supnorm_dist} demonstrate this for the Fourier, PuMA and Cellular datasets, where one may see that the bottleneck distances between PDs are bounded above by the $L^\infty$ distances between the underlying brightness functions. Note in particular that, around $ \sigma = 0.5$, we minimize the sup distance between the original and denoised functions.
  We will see that this value of $\sigma$ is also largely successful in minimizing the error for most of our topological measures.    
  For large values of the denoising parameter $\sigma$ the topological features fade away, and the minimizing bijection for computing the bottleneck distance is to pair every generator with the diagonal (see   $i=0,2$ in Fig.~\ref{fig:fourier_supnorm_dist} for $\sigma \geq 2$). Correspondingly the normalized error here will be precisely $1$ (see Fig.~\ref{fig:fourier_bottleneck_dist}).
  
 We remark that what we look for in a robust topological measure is not a precise optimal value of $\sigma$, but rather we seek to determine measures that possess a reasonably broad range of $\sigma$ for which the denoised dataset is sufficiently close to the original dataset. 
 The coincidence of $ \sigma=0.5$ working well for many measures can be attributed to our using the same strength of noise for all images. Note, moreover, that in an experimental context one would not have access to the ``noise free'' data needed to optimize $\sigma$.

\section{Results} \label{Results}

  We now present our topological analysis of the Fourier, PuMA and Cellular datasets described in Sec.~\ref{Datasets} using the specific measures introduced in Sec.~\ref{Persistence Tools}, appropriately normalized. The results illustrate the distinct topological features of each dataset, and how denoising impacts such features across multiple scales. In each case we present results for the dimension $i=0,1,2$ PH. We first present our results for the Gaussian denoising method in Sec. \ref{sec:gaussian}, and then for the ML-based approach in Sec. \ref{sec:ML}. 

\subsection{Gaussian Denoising\label{sec:gaussian}}

This section discusses results showing how selected topological measures on the noisy and denoised datasets vary with the level of Gaussian denoising, represented by the parameter $\sigma$ in Eq.~\eqref{Gaussian}. The subsections below describe results for each of the six measures introduced in Sec. \ref{Persistence Tools}. 

\subsubsection{Number of Generators}\label{sec:numbergenerators}

\begin{figure}[t]
\centering

\begin{subfigure}[b] {0.32\linewidth}
\centering
\includegraphics[width=\linewidth]{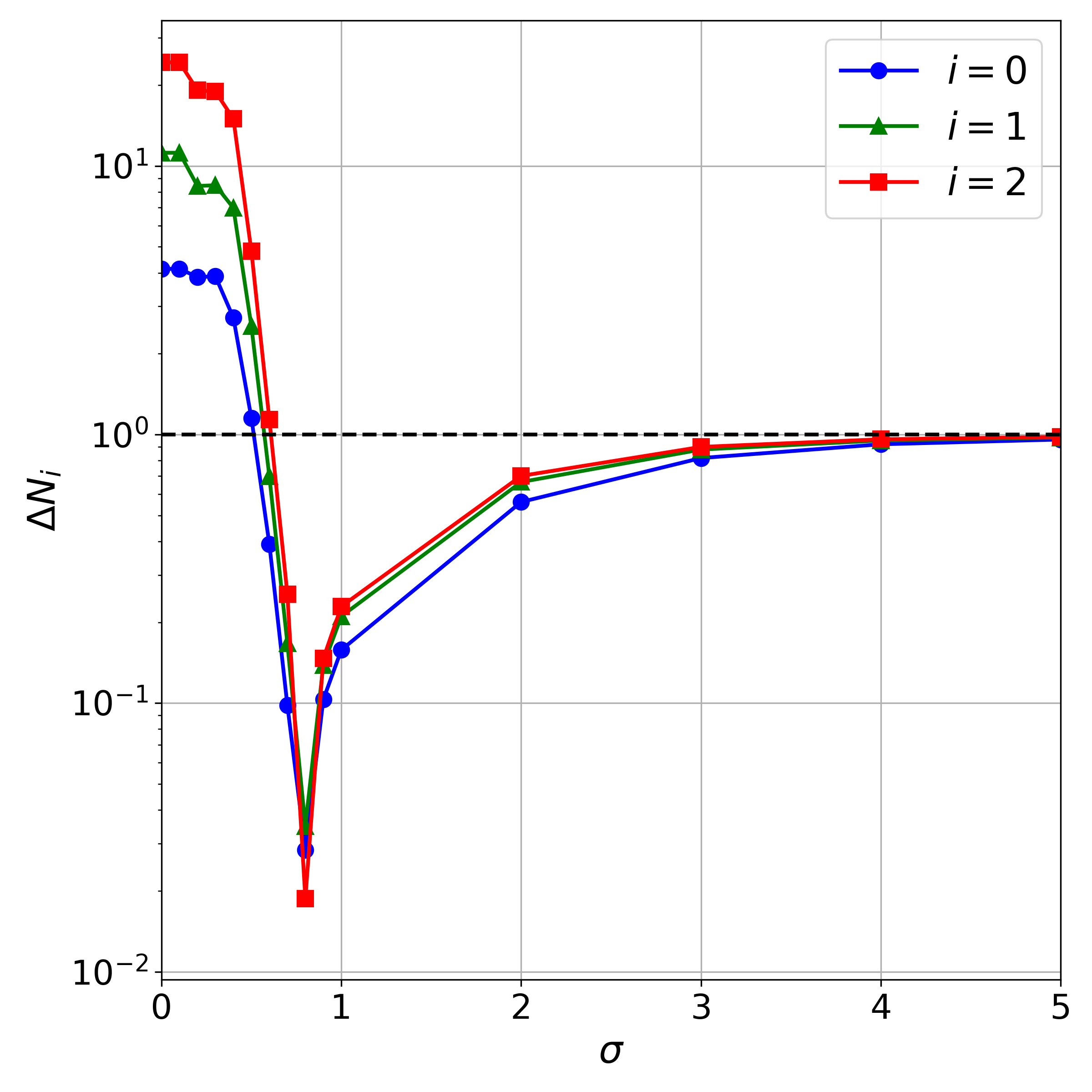}
\caption{\label{fig:fourier_num_generators} Fourier dataset}
\end{subfigure}
\hfill
\begin{subfigure}[b] {0.32\linewidth}
\centering
\includegraphics[width=\linewidth]{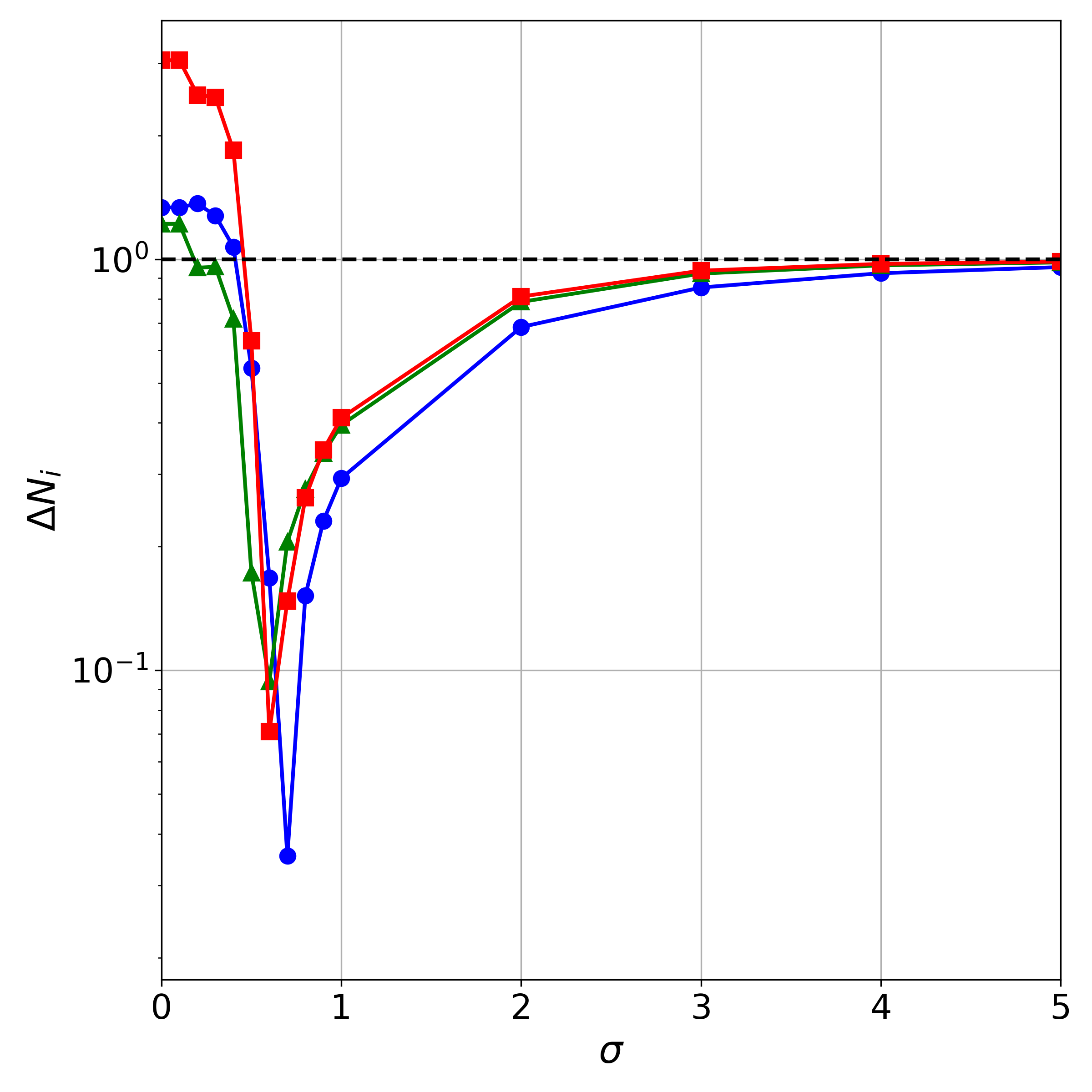}
\caption{\label{fig:puma_norm_num_gen} PuMA dataset}
\end{subfigure}
\hfill
\begin{subfigure}[b] {0.32\linewidth}
\centering
\includegraphics[width=\linewidth]{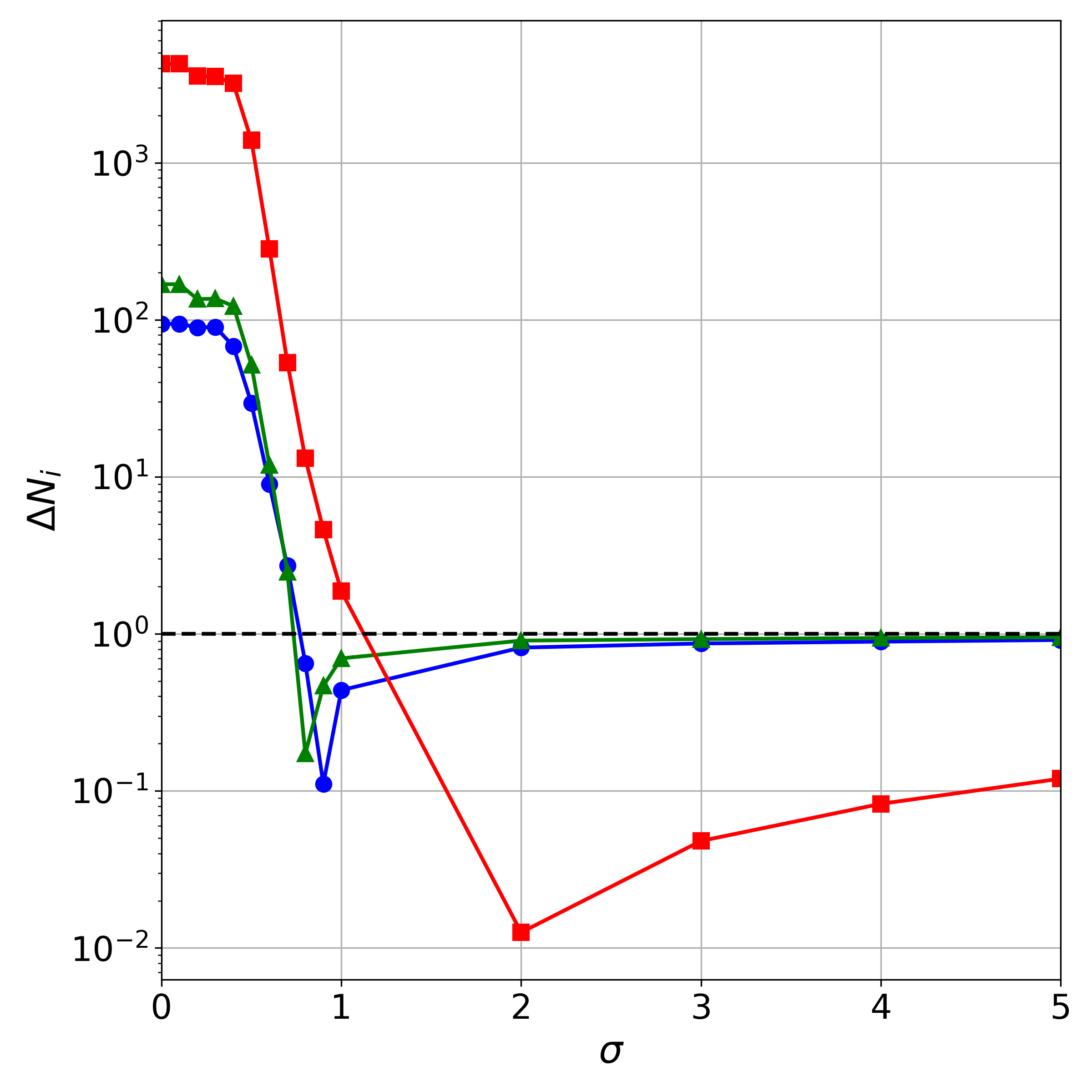}
\caption{\label{cellular_num_generators} Cellular dataset}
\end{subfigure}

\caption{\label{fig:Num_Generators} Normalized difference of the number of generators measure ($\Delta N_i$, see Eq.~(\ref{Number of Generators})) between the original and denoised datasets, plotted on a log scale for easier visualization, for various denoising levels $\sigma \in [0, 5]$.}
\end{figure}

As we saw in the persistence barcodes of Fig.~\ref{fig:fourier_barcode_all}, introducing experimental noise to our datasets leads to a sharp increase in the number of short lived topological features. We also observe this trend in Table \ref{Table: Number of Generators} when comparing the number of generators for each of our datasets before and after the addition of noise. The natural next step is to study how the number of generators of each type varies under the denoising protocol. In Fig.~\ref{fig:Num_Generators} we plot the number of generators error measure, $\Delta N_i:=|N_i^o-N_i^d|/N_i^o$ introduced in Eq.~(\ref{Number of Generators}), representing the normalized difference in the number of generators between the original and denoised datasets, for $i=0,1,2$ and for all three datasets as the denoising parameter $\sigma$ varies. Observe that for $\sigma = 0$, which represents the normalized difference measure between the original and noisy dataset, we have a large error, as anticipated. This is especially true of the Cellular dataset, which originally contains large void spaces that, when noise is introduced, become populated with numerous short-lived features. As $\sigma$ increases, the overall error between the original and denoised datasets decreases. In most cases, we are able to minimize $\Delta N_i$ for $i=0,1,2$ 
  with an optimal denoising value of $\sigma \in [0.5, 0.7]$. The only exception is for $i=2$ for the cellular dataset, which has a much larger optimal denoising value of $\sigma \approx 2$. If we continue to increase $\sigma$ beyond this value, we find that the error increases again, and finally converges to 1 as expected from our normalization; as $\sigma\to\infty$ the denoised images become featureless and $N_i^d\to 0$. 

\begin{figure}[t]
\centering

\begin{subfigure}[b] {0.32\linewidth}
\centering
\includegraphics[width=\linewidth]{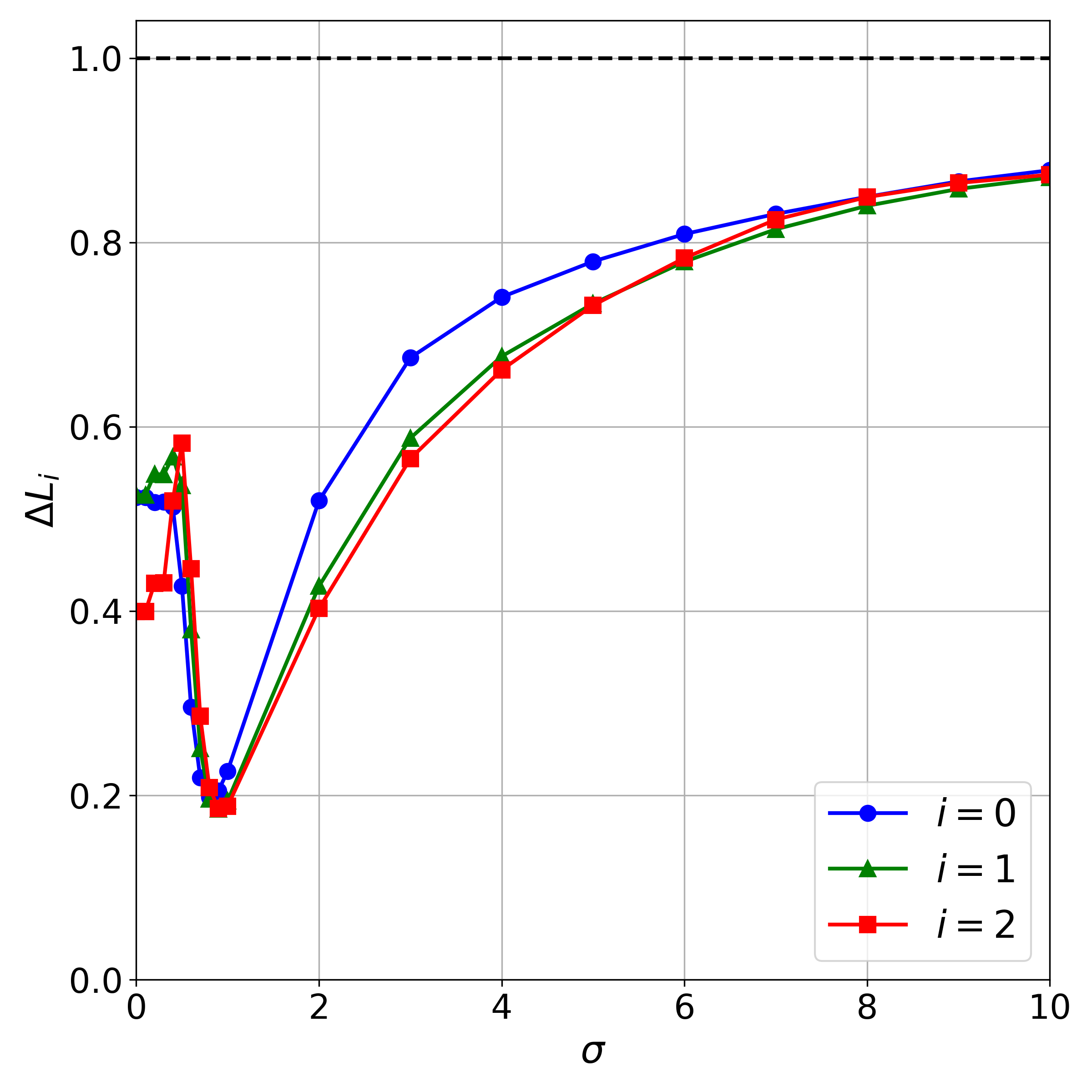}
\caption{\label{fig:fourier_avg_lifespan} Fourier dataset}
\end{subfigure}
\hfill
\begin{subfigure}[b] {0.32\linewidth}
\centering
\includegraphics[width=\linewidth]{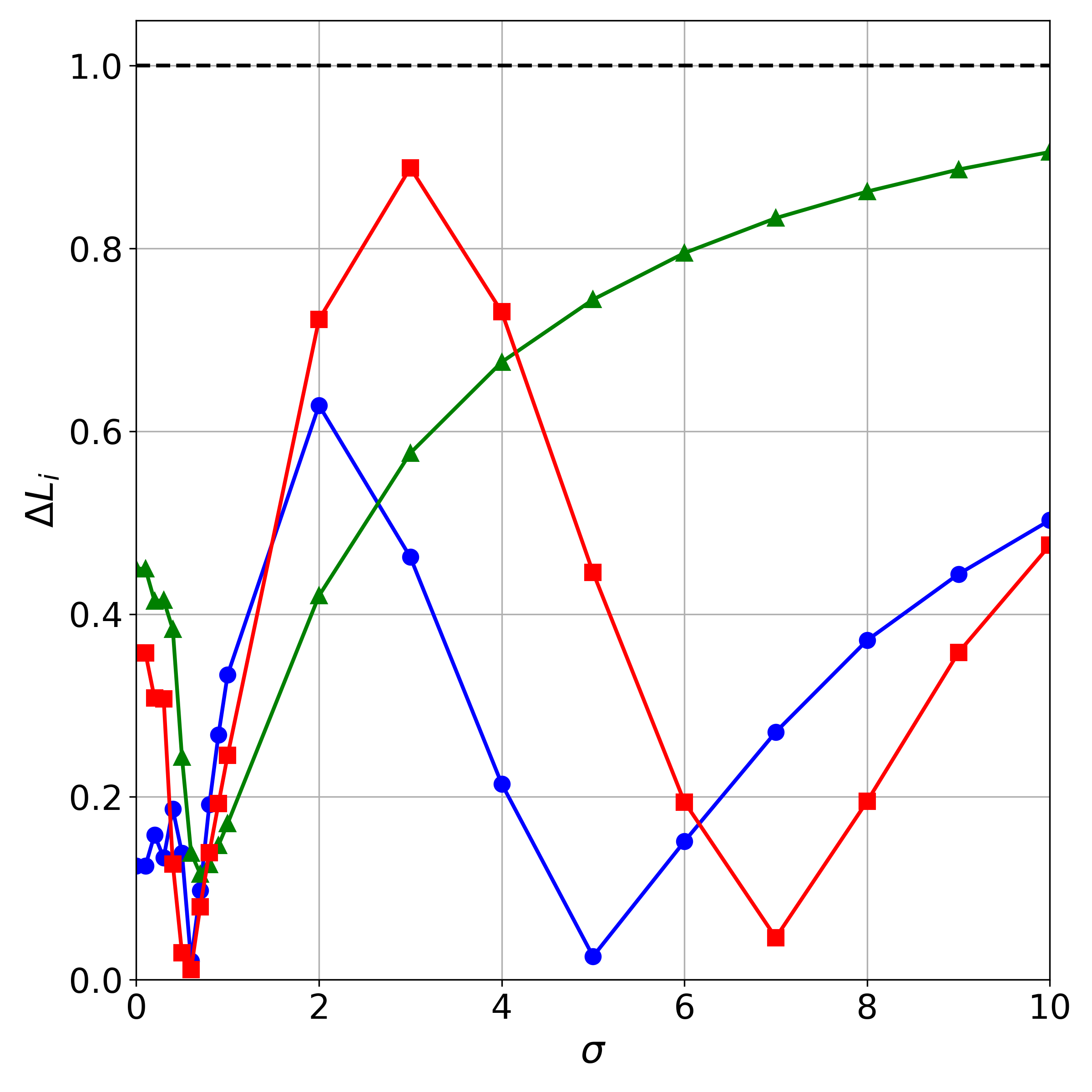}
\caption{\label{fig:puma_norm_av_lifespan} PuMA dataset}
\end{subfigure}
\hfill
\begin{subfigure}[b] {0.32\linewidth}
\centering
\includegraphics[width=\linewidth]{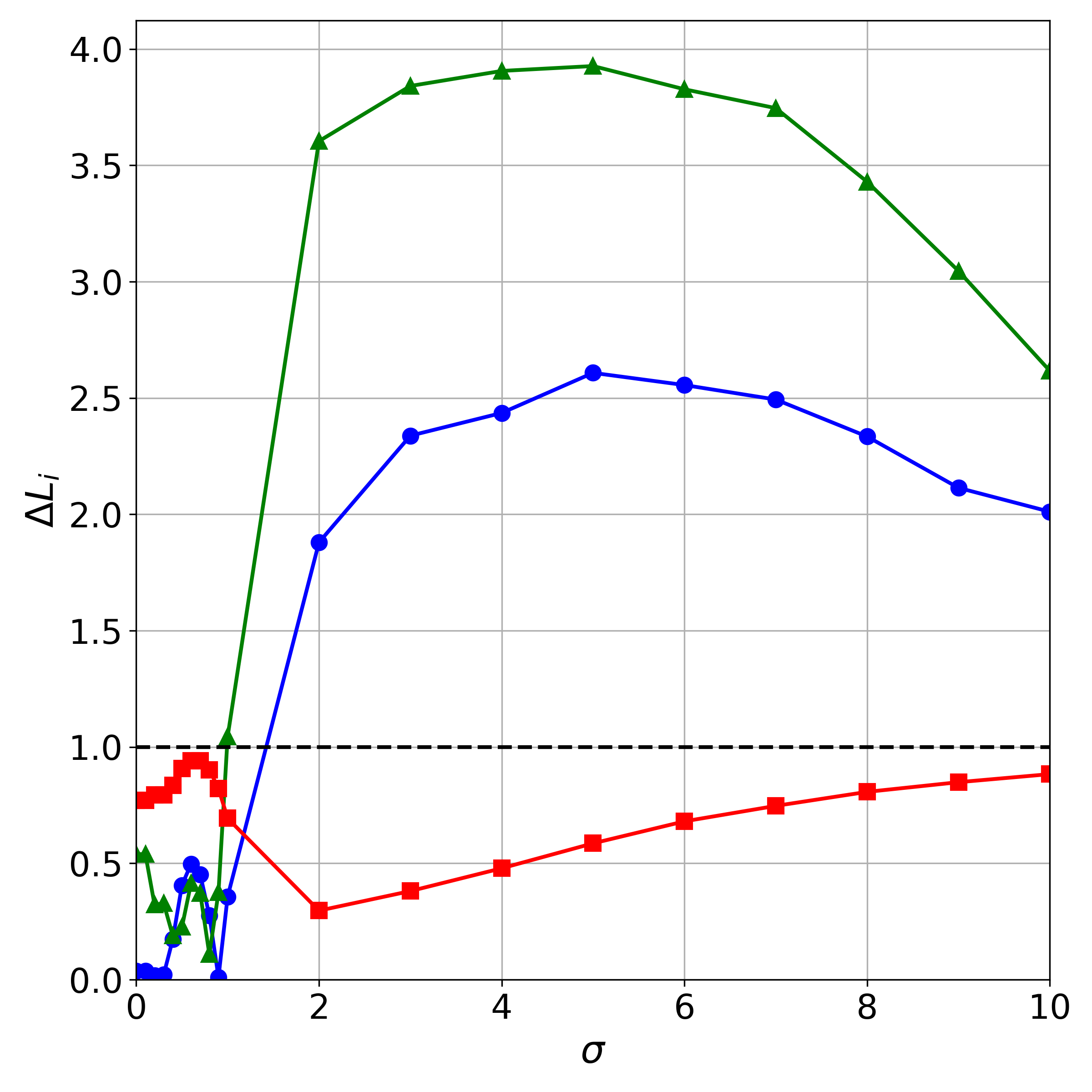}
\caption{\label{cellular_norm_avg_lifespan} Cellular dataset}
\end{subfigure}

\caption{Normalized difference  of the average lifespan measure ($\Delta L_i$, see Eq.~(\ref{Average Lifespan of Generators Measure})) between the original and denoised  datasets, for various denoising levels $\sigma \in [0, 10]$. }
\label{fig:Average_Lifespan}
\end{figure}

\subsubsection{Average Lifespan of Generators}\label{sec:averagelifespan}

  When examining the average lifespan of generators measure~(see Eq.~(\ref{Average Lifespan of Generators Measure})), we observe differing trends across our datasets. In Fig.~\ref{fig:Average_Lifespan} we plot the average lifetime error measure, $\Delta L_i:=|\bar{L}_i^o-\bar{L}_i^d|/\bar{L}_i^o$, representing the normalized difference in average generator lifetimes between original and denoised datasets, versus the smoothing parameter $\sigma$ for $i=0,1,2$ and for all three datasets.  For the Fourier dataset, Fig.~\ref{fig:fourier_avg_lifespan}, we find qualitatively the same behavior as for the number of generators measure discussed above: the error before denoising is relatively high, but reaches a minimum as we increment our denoising parameter within the range $\sigma \in [0.5, 0.7]$. As we continue increasing $\sigma$ beyond the optimal value, we find that the error monotonically asymptotes to 1, reflecting the chosen normalization. 

For the PuMA dataset, Fig.~\ref{fig:puma_norm_av_lifespan}, we observe somewhat different behavior. In line with our previous results, we find optimal values of $\sigma \in [0.5, 0.7]$ that minimize $\Delta L_i$ for $i=0,1,2$. However, we also observe some oscillatory behavior for larger $\sigma$-values, where additional local minima are achieved for the $i=0$ ($\sigma \approx 5$) and $i=2$ ($\sigma \approx 7$) measures. This results from the average lifespan of generators for the denoised dataset unpredictably increasing above and decreasing below the average lifespan of the generators in the original dataset as we increase the value of $\sigma$. A less pronounced though similar trend can also be seen in Fig.~\ref{cellular_norm_avg_lifespan} where we examine the same measure for the Cellular dataset. Ultimately the average lifespan measure must converge to 1 as $\sigma\to\infty$, but here the approach to 1 is not clearly observed within the range studied. These findings lead us to conclude that this measure is not as robust as others for evaluating our denoising method.  

\subsubsection{Bottleneck Distance}\label{sec:bottleneck}

\begin{figure}[t]
\centering
\begin{subfigure}[b] {0.32\linewidth}
\centering
\includegraphics[width=\linewidth]{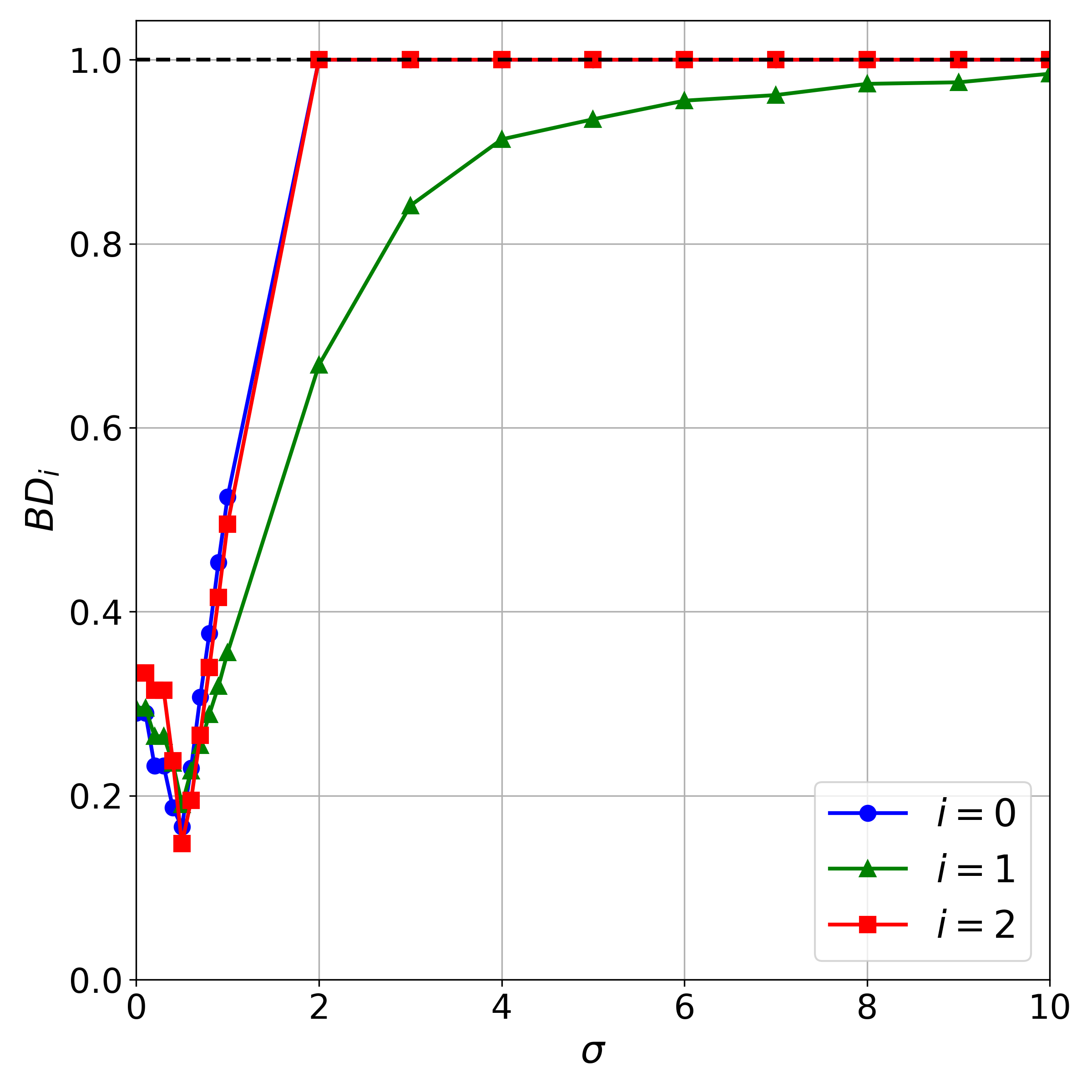}
\caption{\label{fig:fourier_bottleneck_dist} Fourier dataset}
\end{subfigure}
\hfill
\begin{subfigure}[b] {0.32\linewidth}
\centering
\includegraphics[width=\linewidth]{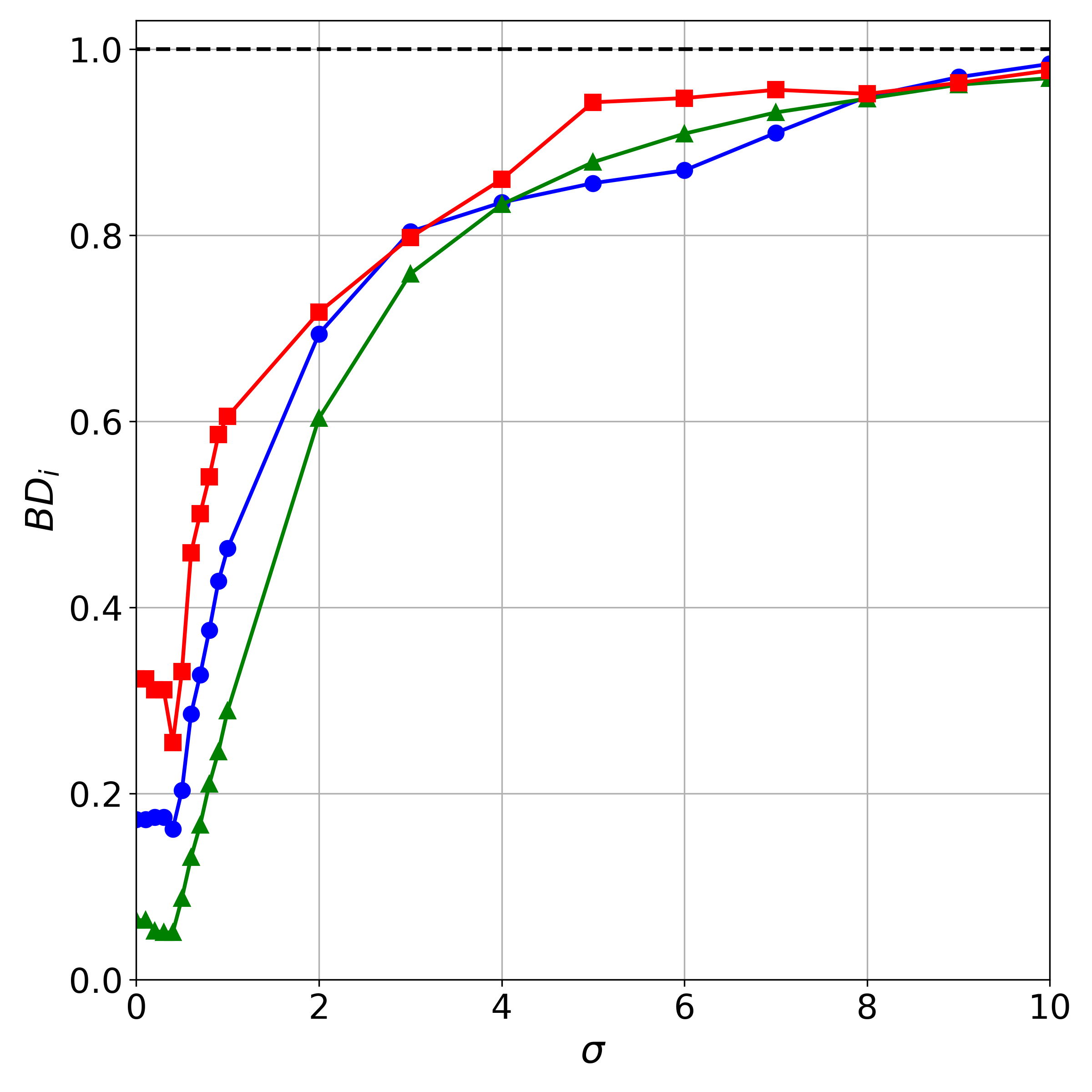}
\caption{\label{fig:puma_norm_bottleneck_dist} PuMA dataset}
\end{subfigure}
\hfill
\begin{subfigure}[b] {0.32\linewidth}
\centering
\includegraphics[width=\linewidth]{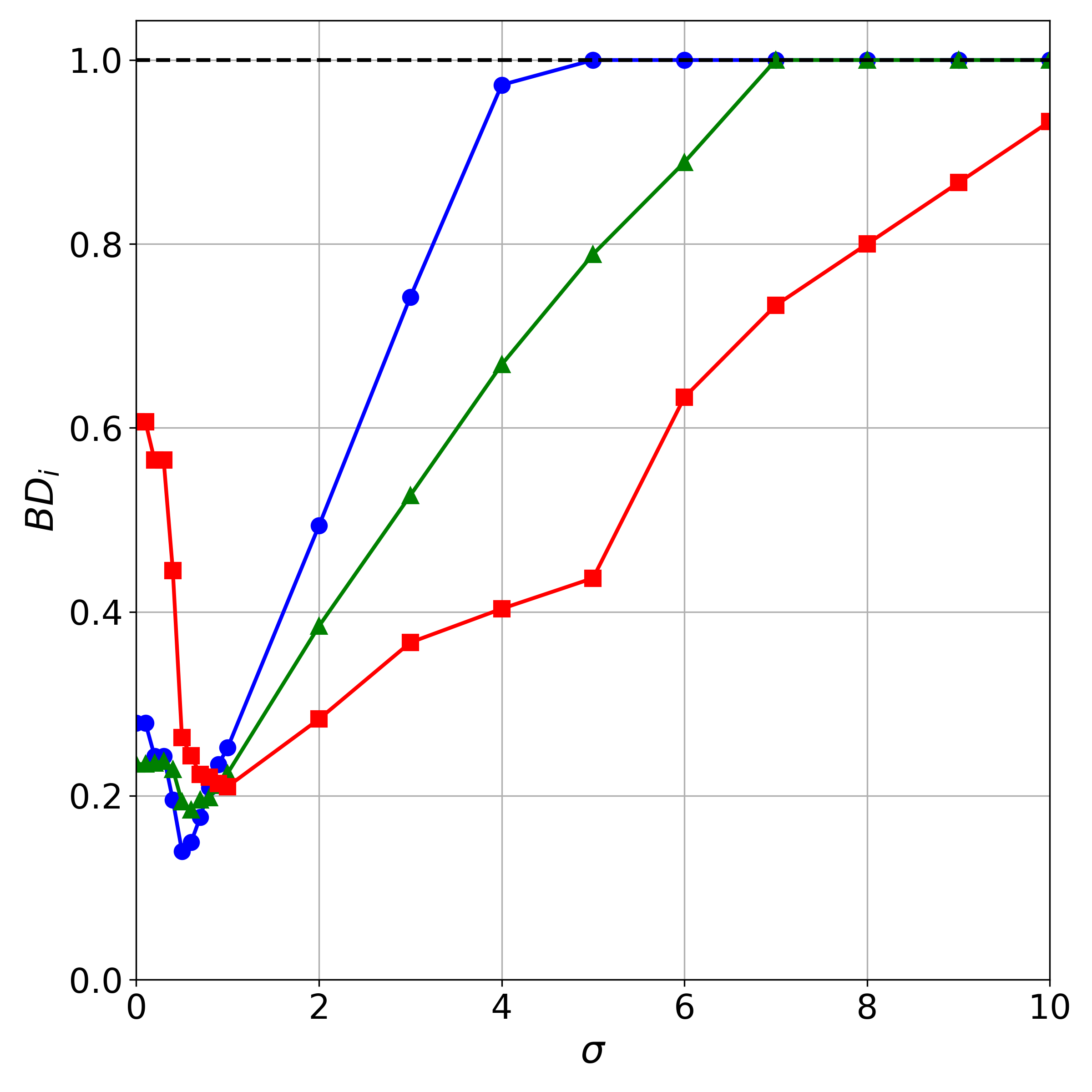}
\caption{\label{cellular_bottleneck_dist} Cellular dataset}
\end{subfigure}
\caption{\label{fig:Bottleneck} Normalized  bottleneck distance  ($BD_i$, see Sec.~\ref{sec:bottleneck}) between the original and denoised datasets, for various denoising levels $\sigma \in [0, 10]$. }
\end{figure}

 In Fig.~\ref{fig:Bottleneck} we present results for the bottleneck distance measure based on $d_b$~(see Definition \ref{Bottleneck Distance}) across various levels of denoising for our three datasets. Specifically, we plot $BD_i :=d_b({\rm PD}_i^o, {\rm PD}_i^d)/d_b({\rm PD}^o_i,\emptyset)$, the bottleneck distance between the denoised PD and the original PD associated with the $i$-th homology group of these datasets, normalized by the bottleneck distance between the original PD and the empty PD. A point of interest is that here the initial error before denoising is relatively low compared to the measures discussed so far. This is likely because the bottleneck distance (unlike the Wasserstein distance discussed next) only takes into account the largest distance pairing between the objects compared. As a consequence, the bottleneck distance measure is less sensitive to the sharp increase in the number of non-persistent generators when noise is added to our original datasets. 
 
\begin{figure}[t]
\centering
\begin{subfigure}[b] {0.32\linewidth}
\centering
\includegraphics[width=\linewidth]{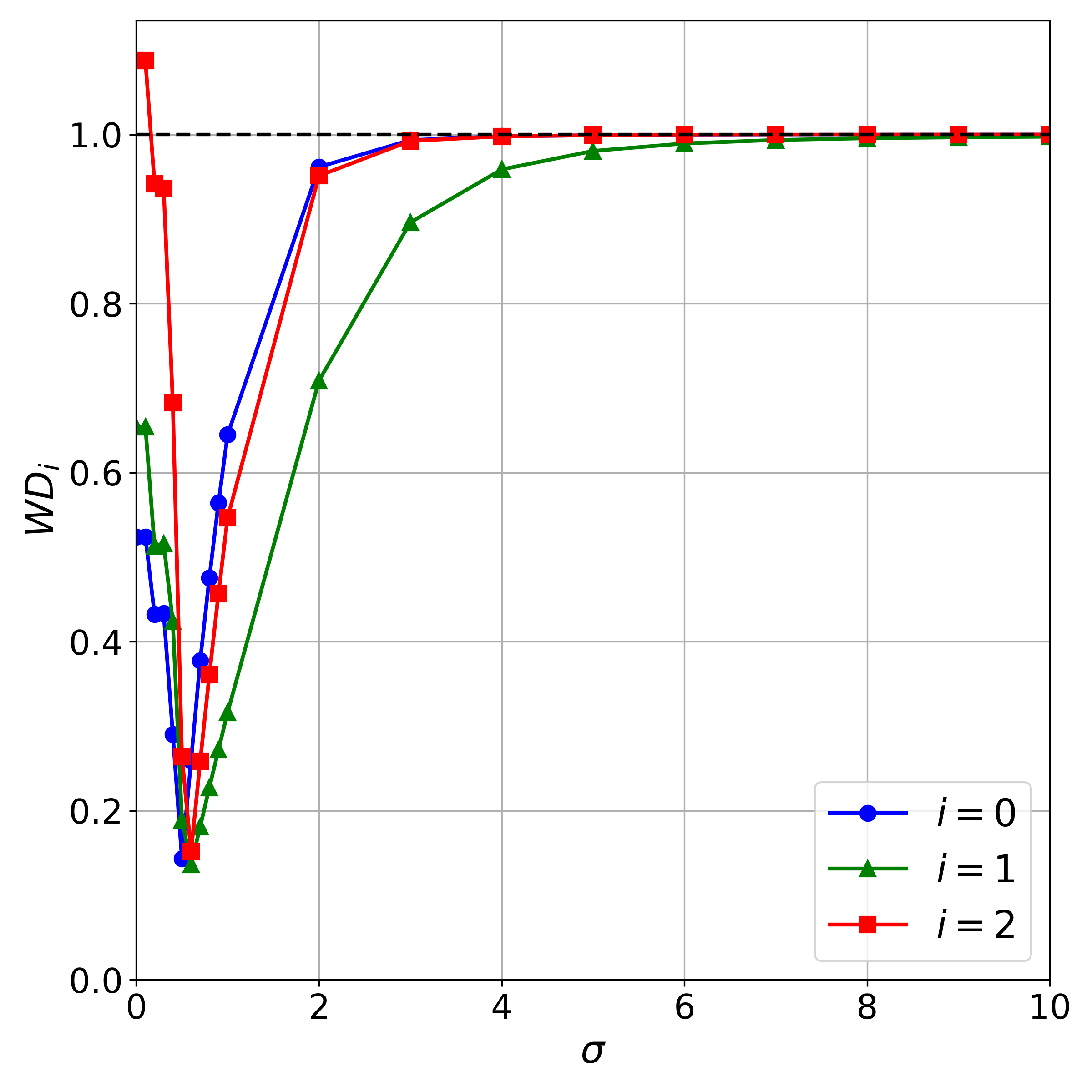}
\caption{\label{fig:fourier_wasser_distance} Fourier dataset}
\end{subfigure}
\hfill
\begin{subfigure}[b] {0.32\linewidth}
\centering
\includegraphics[width=\linewidth]{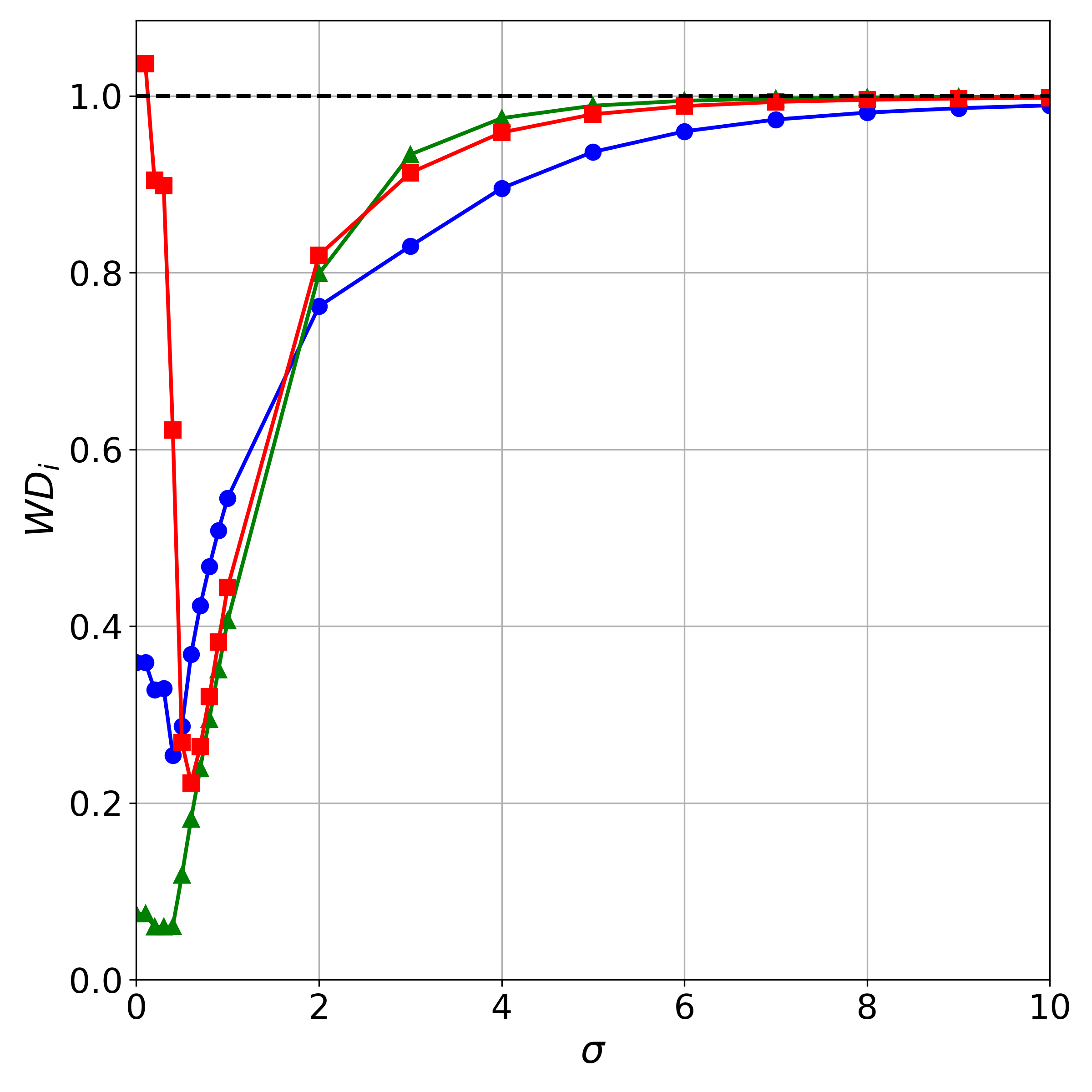}
\caption{\label{fig:puma_norm_wasser} PuMA dataset}
\end{subfigure}
\hfill
\begin{subfigure}[b] {0.32\linewidth}
\centering
\includegraphics[width=\linewidth]{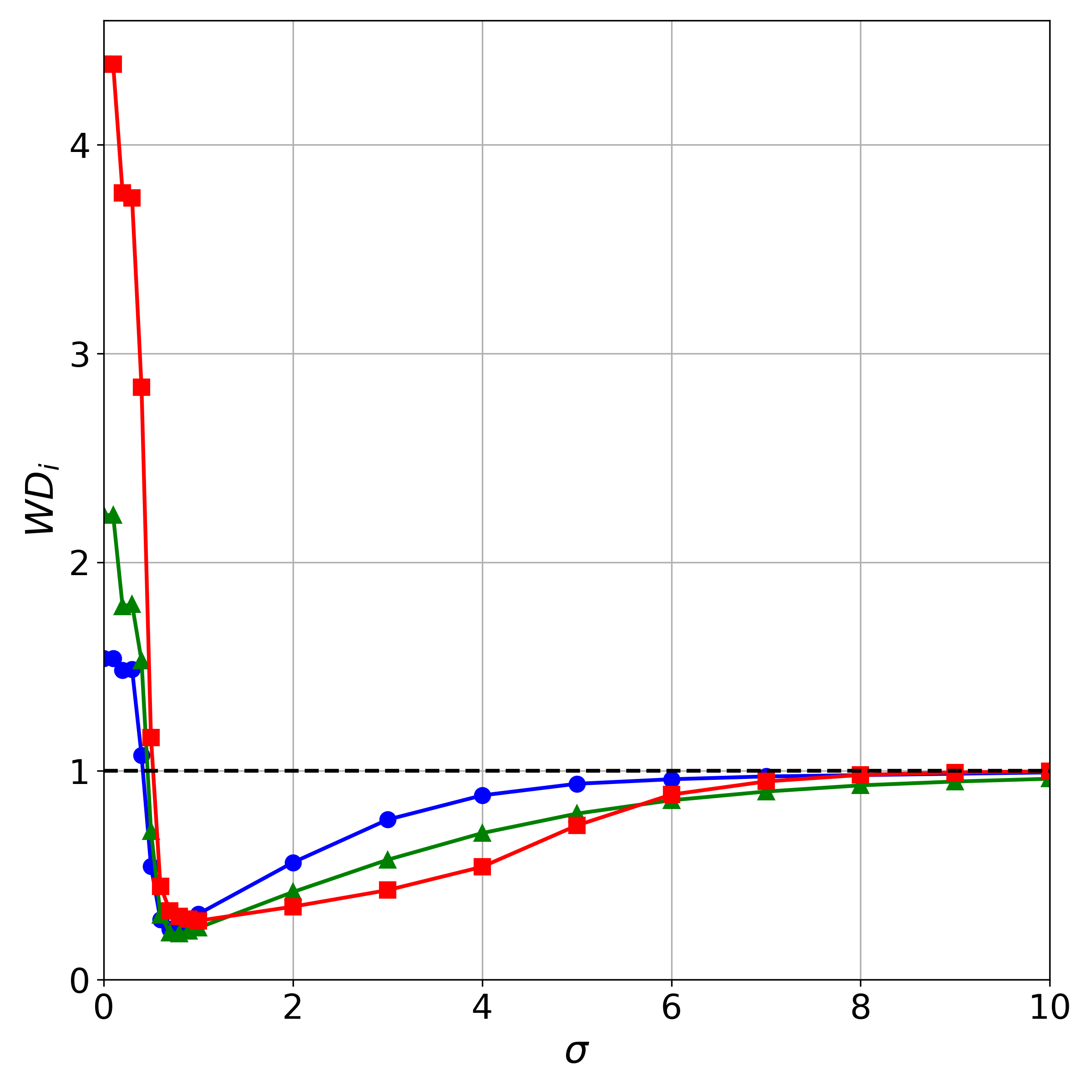}
\caption{\label{cellular_norm_wasser} Cellular dataset}
\end{subfigure}
\caption{\label{fig:Wasserstein} Normalized Wasserstein distance  ($WD_i$ for $i=0,1,2$, see Sec.~\ref{sec:wasserstein}) between the original and denoised  datasets, for various denoising levels $\sigma \in [0, 10]$. }
\end{figure}

\subsubsection{Wasserstein Distance}\label{sec:wasserstein}

 In Fig.~\ref{fig:Wasserstein}, we present the analogous results for the measure based on the Wasserstein distance (see Definition \ref{Wasserstein Distance}), $WD_i:=W_2({\rm PD}_i^o, {\rm PD}_i^d)/W_2({\rm PD}^o_i,\emptyset)$, for $i=0,1,2$, as a function of $\sigma$ for all three of our datasets. The measure behaves as anticipated for an effective denoising method: the error is relatively high before denoising, but is minimized for some optimal $\sigma \in [0.5, 0.7]$. As expected from the chosen normalization, the error approaches 1 as we continue increasing $\sigma$. Similar to the number of generators measure, the initial error here may be large. This similarity can be seen in the definition of the Wasserstein distance, which is sensitive to the large number of non-persistent generators created by introducing random noise into the original dataset. 

\begin{figure}[t]
\centering

\begin{subfigure}[b] {0.32\linewidth}
\centering
\includegraphics[width=\linewidth]{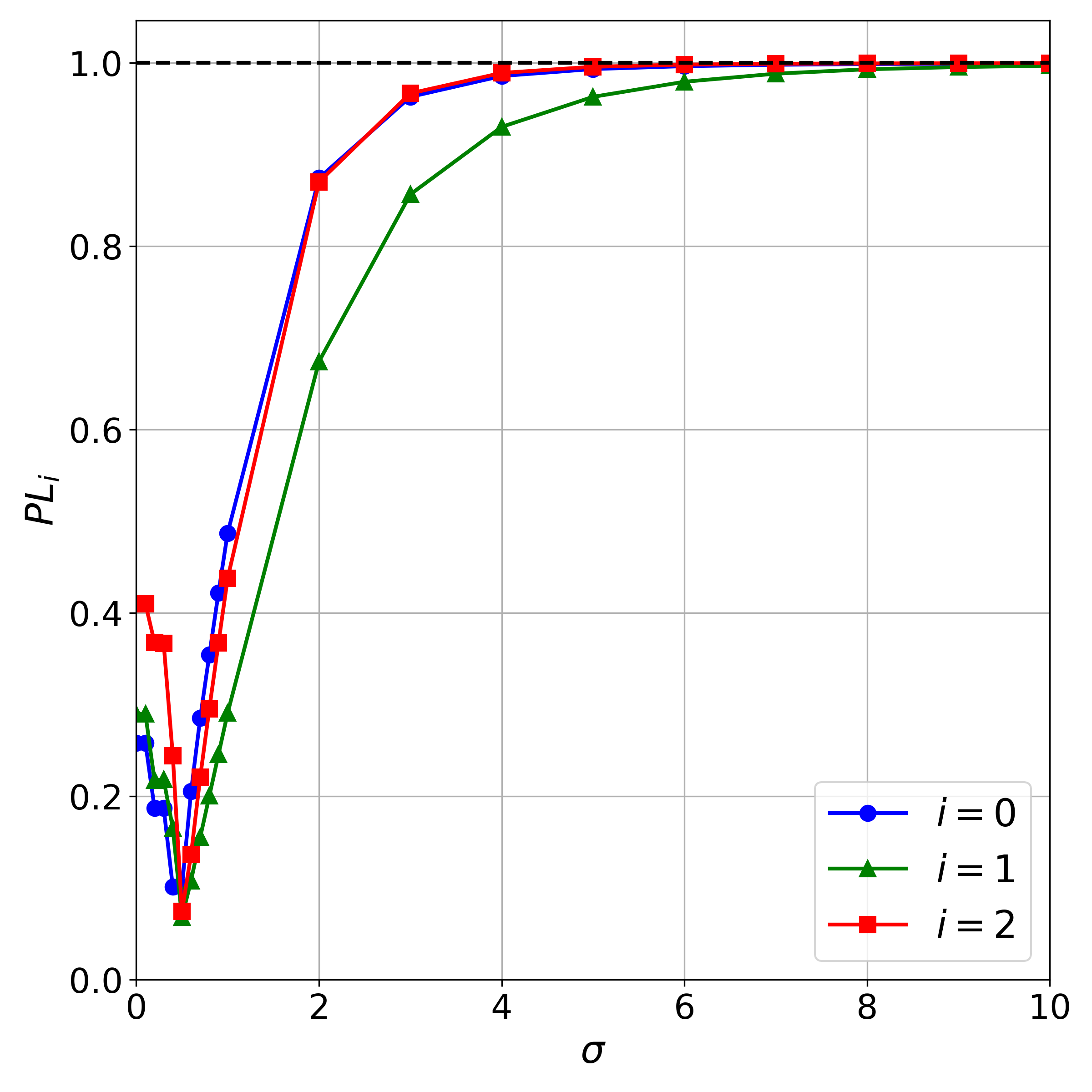}
\caption{\label{fig:fourier_L2_PL100} Fourier dataset}
\end{subfigure}
\hfill
\begin{subfigure}[b] {0.32\linewidth}
\centering
\includegraphics[width=\linewidth]{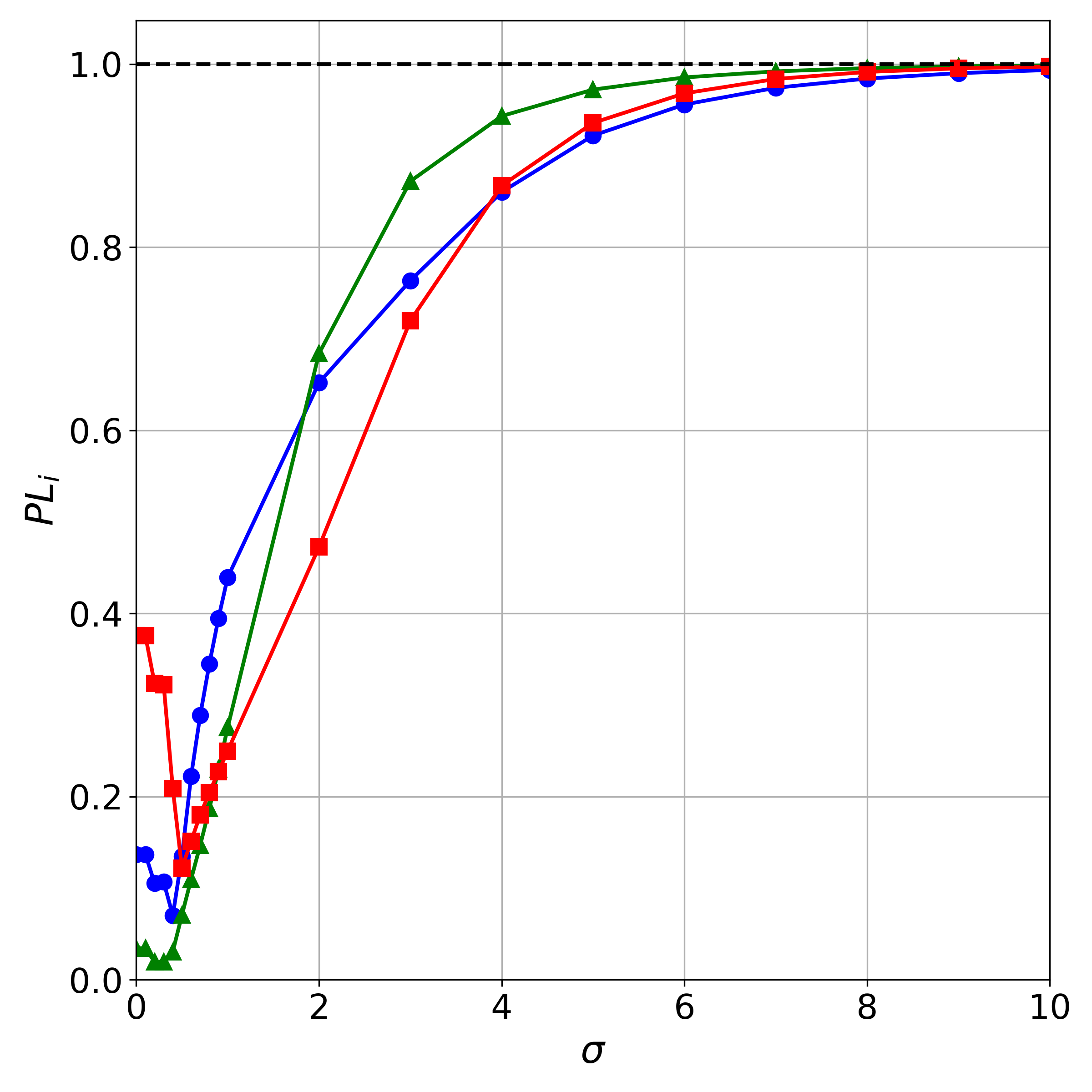}
\caption{\label{fig:puma_L2_PL100} PuMA dataset}
\end{subfigure}
\hfill
\begin{subfigure}[b] {0.32\linewidth}
\centering
\includegraphics[width=\linewidth]{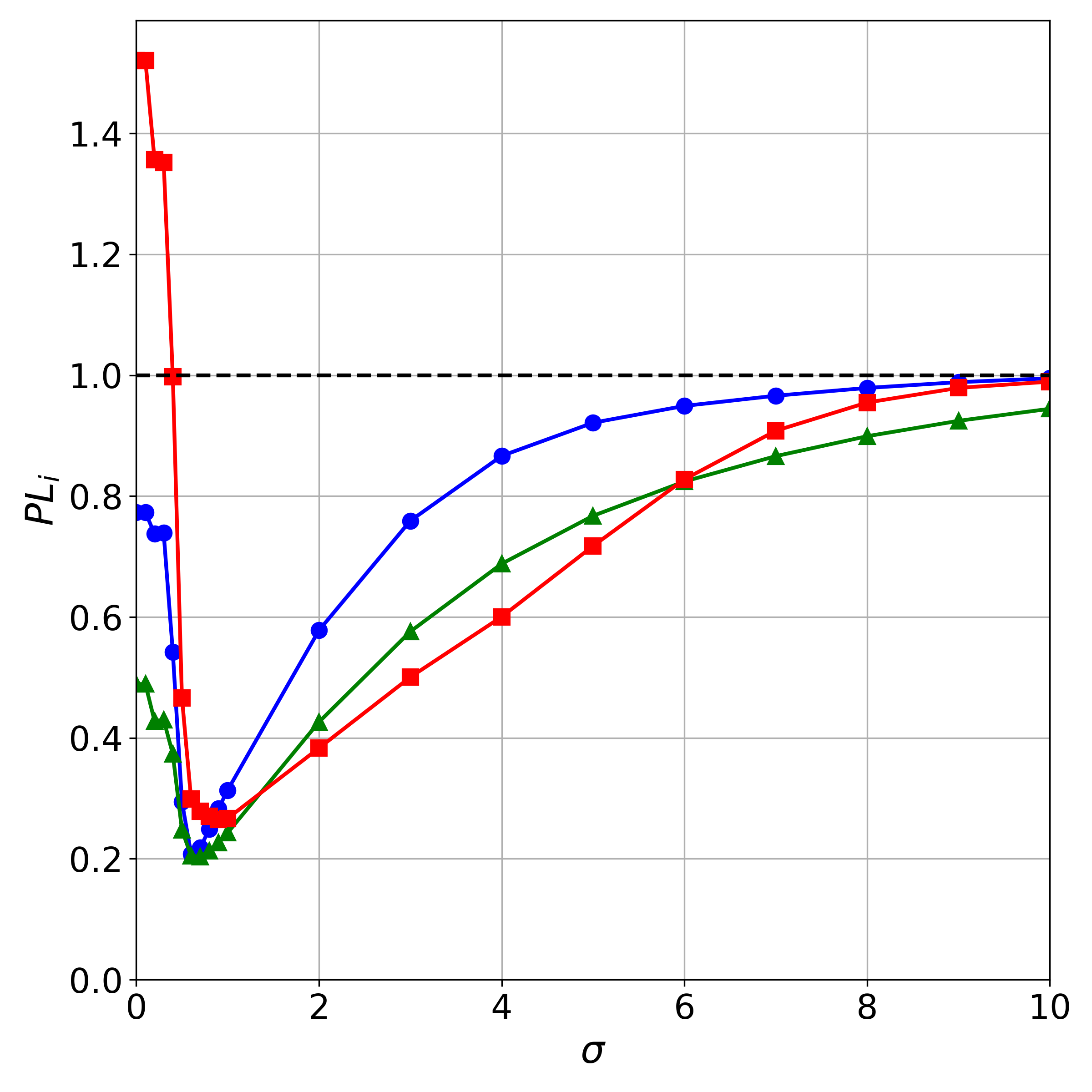}
\caption{\label{cellular_L2_PL100} Cellular dataset}
\end{subfigure}

\caption{\label{fig:L2_PL} 
 Normalized  $L^2$ distance, $PL_{i}$ (see Sec.~\ref{sec:PLmeasure}) between  persistence landscapes of the original and denoised datasets, for various denoising levels $\sigma \in [0, 10]$. }
\end{figure}
\begin{figure}[t]
\centering

\begin{subfigure}[b] {0.32\linewidth}
\centering
\includegraphics[width=\linewidth]{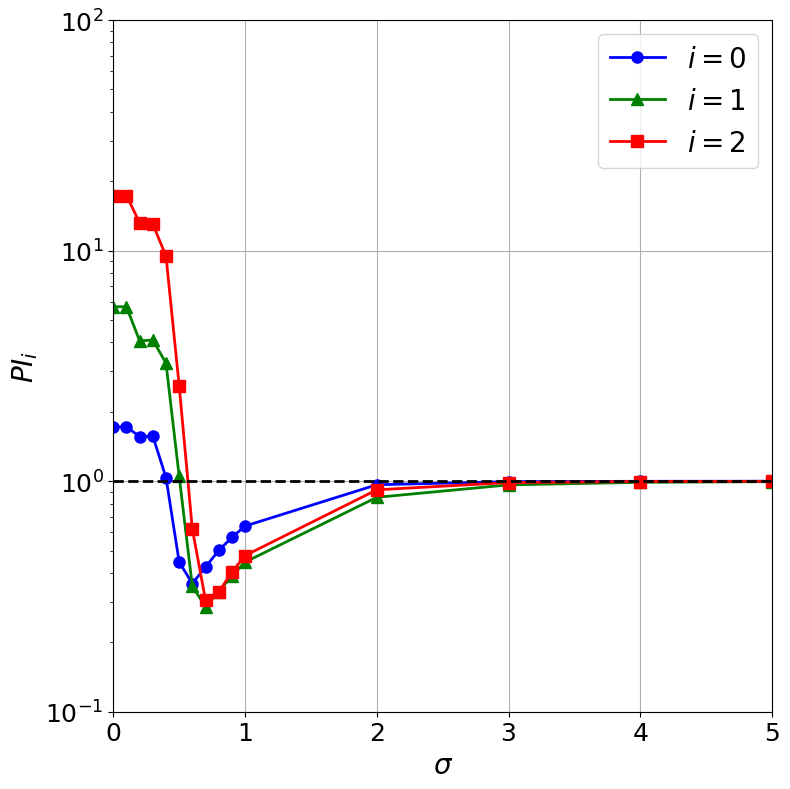}
\caption{\label{fig:fourier_l2_PI} Fourier dataset}
\end{subfigure}
\hfill
\begin{subfigure}[b] {0.32\linewidth}
\centering
\includegraphics[width=\linewidth]{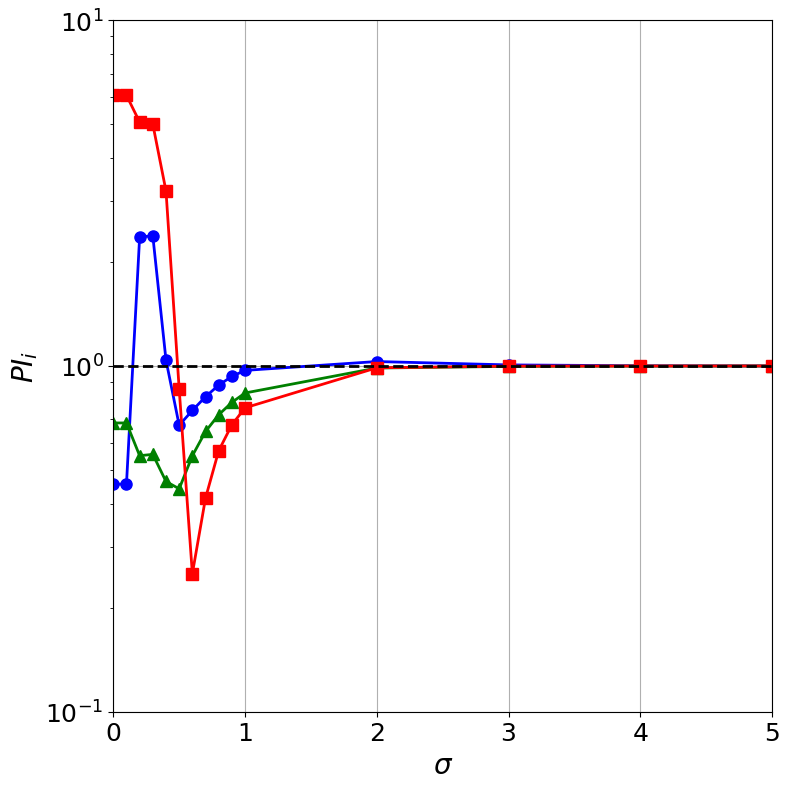}
\caption{\label{fig:puma_l2_PI} PuMA dataset}
\end{subfigure}
\hfill
\begin{subfigure}[b] {0.32\linewidth}
\centering
\includegraphics[width=\linewidth]{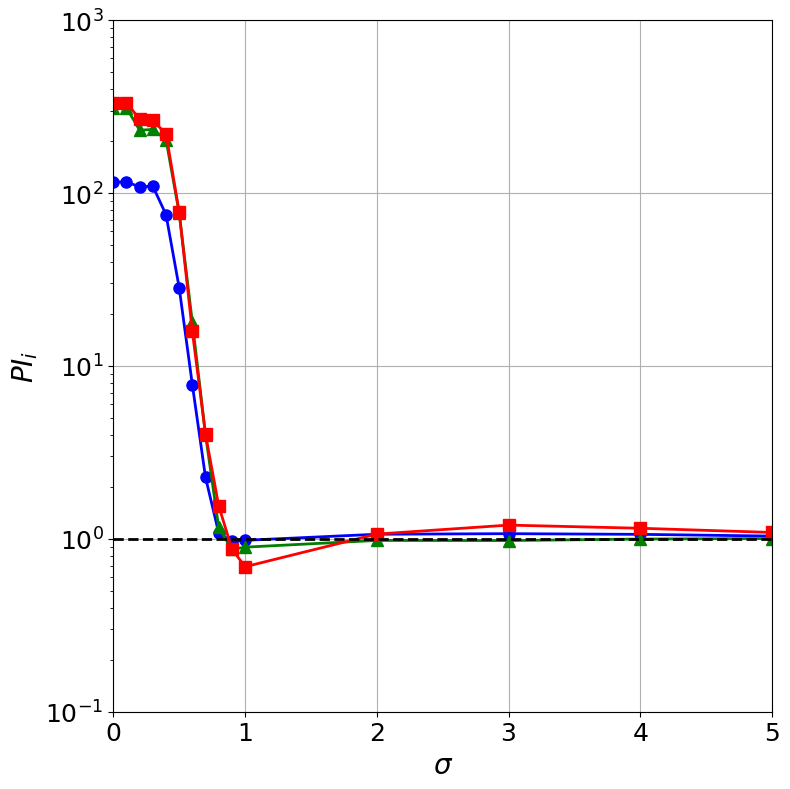}
\caption{\label{cellular_l2_PI} Cellular dataset}
\end{subfigure}

\caption{\label{fig:L2_PI} Normalized $L^2$ distance $PI_{i}$ (see Sec.~\ref{sec:persistenceimage}) between persistence images of the original and denoised   datasets, for various denoising levels $\sigma \in [0, 5]$. }
\end{figure}

\subsubsection{Persistence Landscapes}\label{sec:PLmeasure}

In Fig.~\ref{fig:L2_PL}, we present the results for 
\begin{equation}
\label{PL measure}
PL_i:=\|\lambda_{{\rm PD}_i^o} - \lambda_{{\rm PD}_i^d}\|_2/\|\lambda_{{\rm PD}_i^o}\|_2 
\end{equation}
(see Definition \ref{def:Persistence Landscape}), the $L^2$ norm of the difference between the denoised persistence landscape and the original persistence landscape of the datasets, normalized by the $L^2$ norm of the original persistence landscape. As usual we show results for the dimension $i=0,1,2$ PH for all three datasets. To speed up the runtime of this computation (and avoid calculating millions of landscape functions), we only consider the first 100 upper envelopes as defined in Definition~\ref{def:Persistence Landscape} when calculating this measure. In addition, to ensure that the $L^2$ norm is well defined we discretize the original and denoised persistence landscapes with the same meshes. This guarantees that the resulting vectorized representations have an equal number of entries.

 As we vary the denoising parameter $\sigma$, we find that the same pattern seen for our previous measures emerges. Similar to the bottleneck distance, the error before denoising is again relatively low here --- an indication that this measure is also more resilient to noise than the number of generators and Wasserstein distance measures. 

\subsubsection{Persistence Images}\label{sec:persistenceimage}

The last measure we consider is the persistence image error measure, 
\begin{equation}
\label{PI measure}
    PI_{i}:=\|I_{{\rm PD}_i^o} - I_{{\rm PD}_i^d}\|_2 / \|I_{{\rm PD}_i^o}\|_2,
\end{equation} 
the $L^2$ norm between the denoised persistence images and the original persistence images of each of our datasets, normalized by the $L_2$ norm of the original persistence image. As with the persistence landscapes, to ensure that the $L^2$ norm is well-defined for this measure, we discretize the original and denoised persistence images using the same meshes, ensuring that the resulting vectorized representations have the same number of entries. Fig.~\ref{fig:L2_PI} shows this measure $PI_{i}$ for $i=0,1,2$ and for varying values of $\sigma$, for each of our three datasets. We find that the same general trend seen for the other measures emerges (we do observe some minor oscillatory behavior as $\sigma$ varies, but less extreme than that seen for the average lifespan of generators measure in Fig.~\ref{fig:Average_Lifespan}). A curiosity for this measure is that the initial error before denoising is again relatively high, similar to the number of generators and (to a lesser extent) the Wasserstein distance measures. This is surprising given that we are still using the $L^2$ norm to compare the persistence images, as we did in Fig.~\ref{fig:L2_PL}, which by contrast showed initially small errors.

\subsection{Machine Learning (ML) based Denoising\label{sec:ML}}

We now present results for the ML-based denoising approach described in Sec.~\ref{Noising and Denoising}. Following Sec.~\ref{sec:gaussian}, we report on four of the six measures from Sec.~\ref{Persistence Tools}, excluding bottleneck and Wasserstein distance measures due to prohibitive computation times. For each measure, histograms display the distribution of ML model performance across the 90 test datasets with the horizontal axis representing the measure value and the vertical axis representing the frequency of occurrence.

\subsubsection{Number of Generators}\label{sec:MLnumbergenerators}

\begin{figure}[H]
\begin{subfigure}[b] {0.32\linewidth}
\centering
\includegraphics[width=\linewidth]{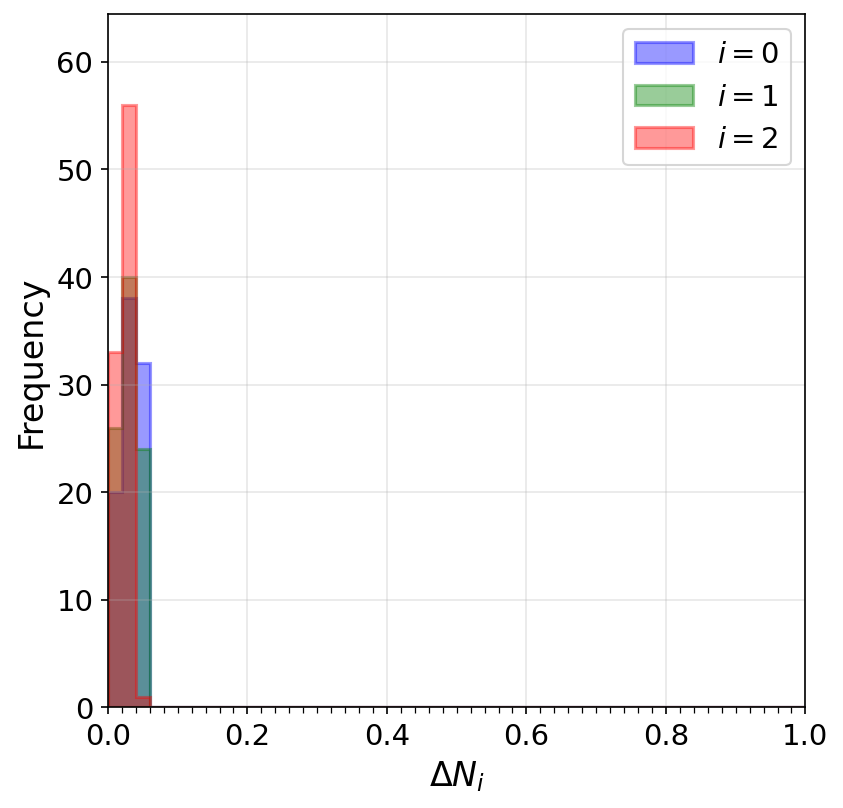}
\caption{\label{fig:fourier_ML_NG} Fourier dataset}
\end{subfigure}
\hfill
\begin{subfigure}[b] {0.32\linewidth}
\centering
\includegraphics[width=\linewidth]{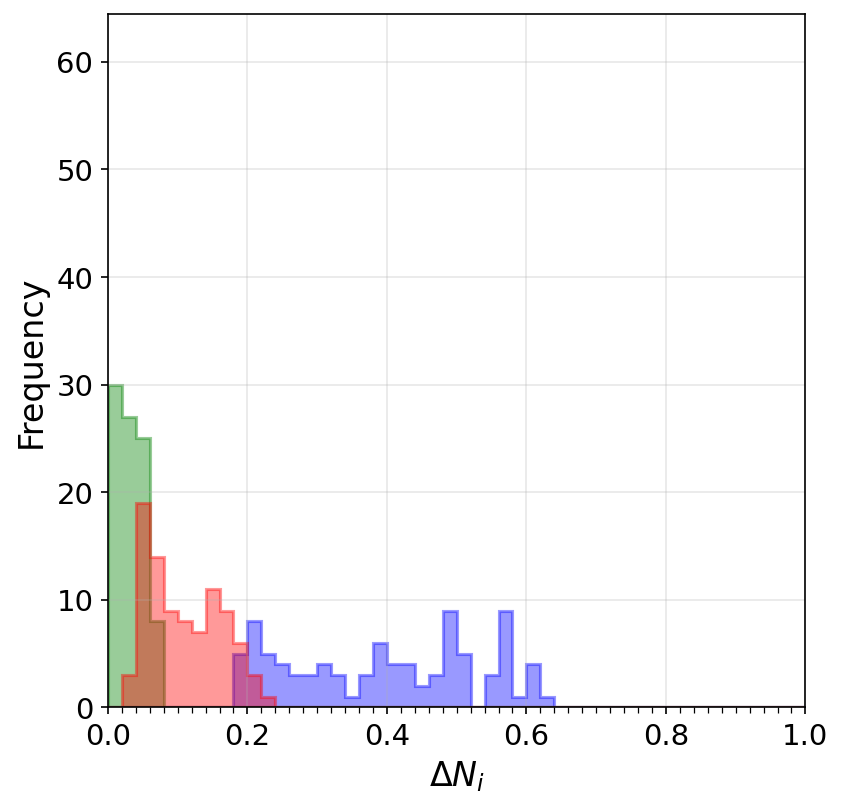}
\caption{\label{fig:puma_ML_NG} PuMA dataset}
\end{subfigure}
\hfill
\begin{subfigure}[b] {0.32\linewidth}
\centering
\includegraphics[width=\linewidth]{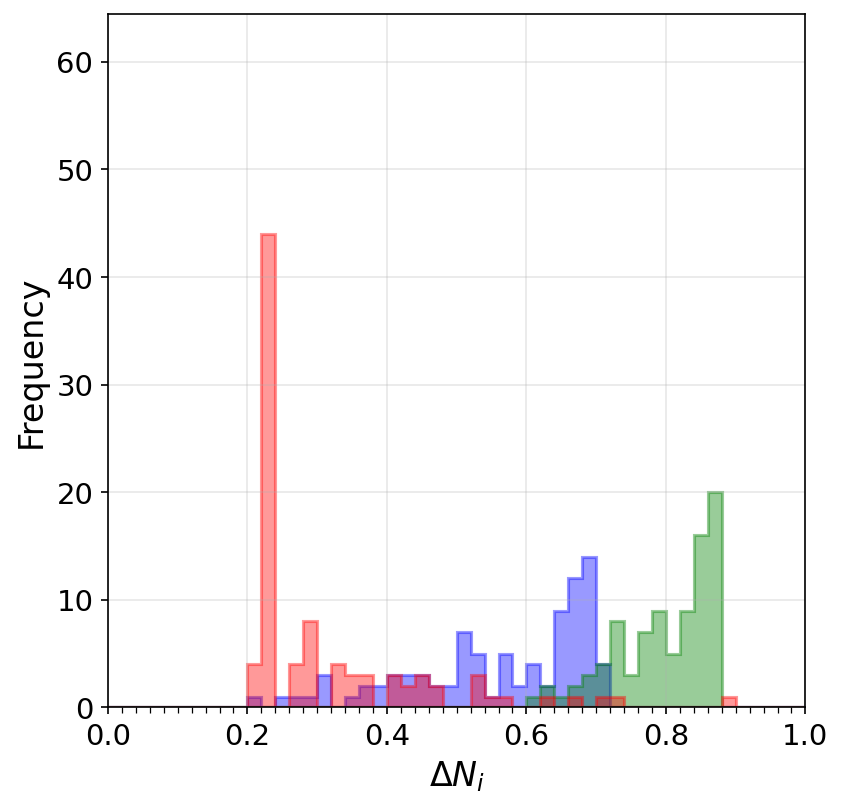}
\caption{\label{fig:cellular_ML_NG} Cellular dataset}
\end{subfigure}

\caption{\label{fig:NG_ML} Distribution of the normalized difference of the number of generators measure ($\Delta N_i$, see Eq.~(\ref{Number of Generators})) over $90 \times 3$ testing datasets.}
\end{figure}

Figure~\ref{fig:NG_ML} presents the distribution of the number of generators measure $\Delta N_i$, defined in Eq.~\eqref{Number of Generators}, across the 90 testing images for each of the morphological datasets for homological dimensions $i=0, 1,2$. We can observe from Fig.~\ref{fig:fourier_ML_NG} that for the Fourier dataset, the measure exhibits distributions tightly concentrated near zero for all three dimensions, indicating that the U-Net largely preserves the number of connected components, loops, and voids present in the original images. This behavior, however, does not reliably generalize across datasets. For instance, in Fig.~\ref{fig:puma_ML_NG} we see that the distributions for $i = 1$ and $i = 2$ remain relatively concentrated, yet exhibit noticeably heavier tails, suggesting that the denoising model occasionally introduces or destroys a non-trivial number of topological features. The measure most sharply deviates from the desired behavior for the cellular dataset, as seen in Fig.~\ref{fig:cellular_ML_NG}, where the distributions for $i = 0$ and $i = 1$ are broad and spread across nearly the full range $[0, 1]$, while the $i = 2$ distribution is quite sharply concentrated near $\Delta N_i \approx 0.2$--$0.3$. This dataset-dependent variability reveals that $\Delta N_i$ is a highly sensitive topological measure where small structural differences between training and test distributions can produce large fluctuations in the number of generators, making this measure an unreliable indicator of overall denoising quality. 

\subsubsection{Average Lifespan of Generators}\label{sec:MLaveragelifespan}

\begin{figure}[H]
\begin{subfigure}[b] {0.32\linewidth}
\centering
\includegraphics[width=\linewidth]{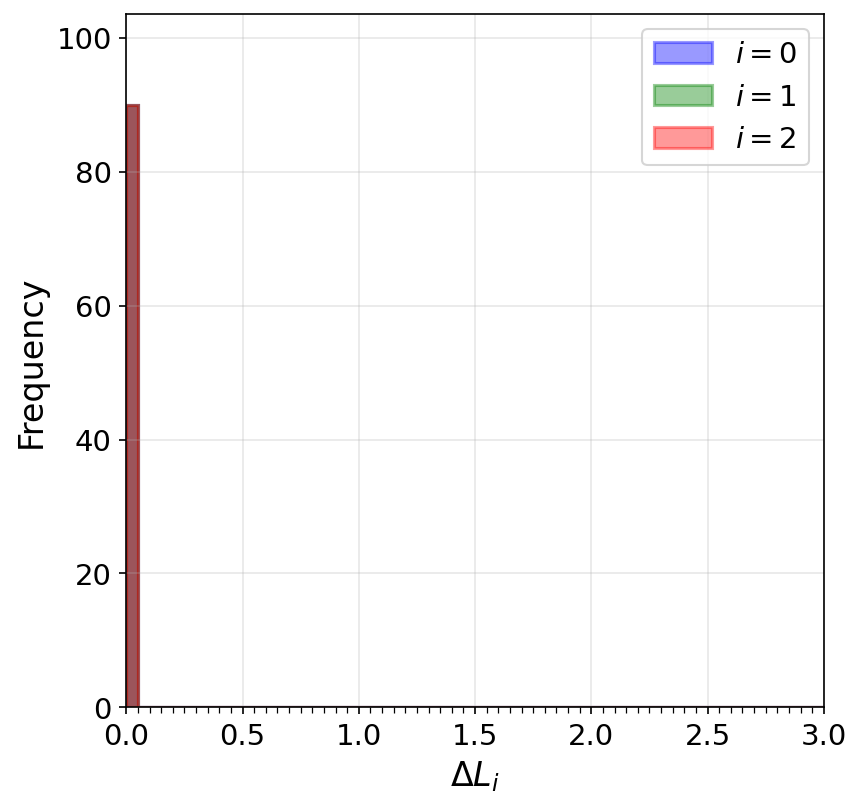}
\caption{\label{fig:fourier_ML_AL} Fourier dataset}
\end{subfigure}
\hfill
\begin{subfigure}[b] {0.32\linewidth}
\centering
\includegraphics[width=\linewidth]{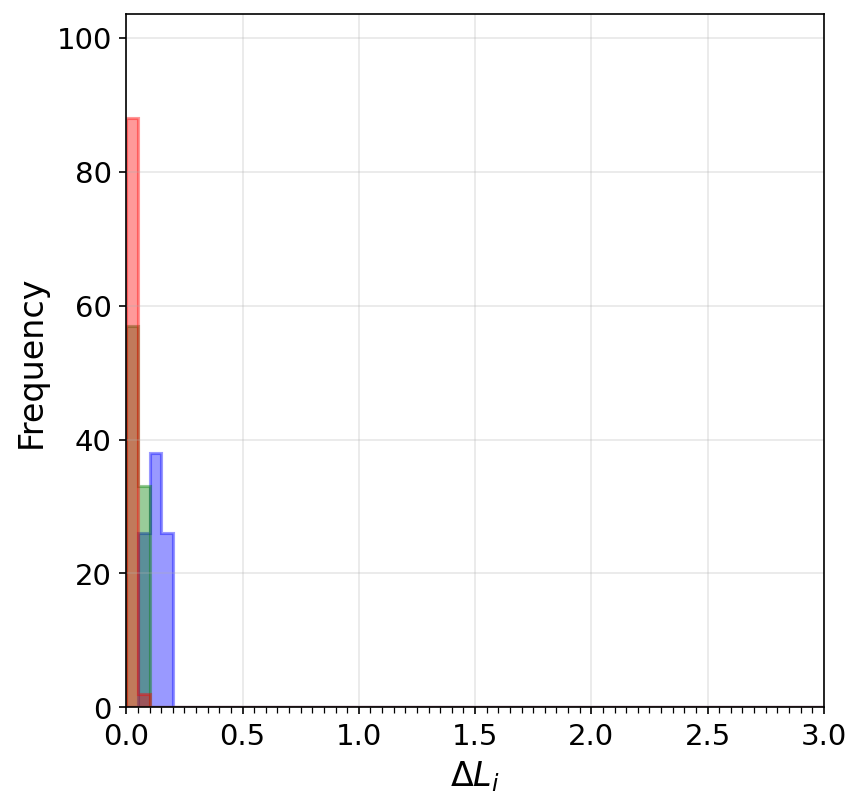}
\caption{\label{fig:puma_ML_AL} PuMA dataset}
\end{subfigure}
\hfill
\begin{subfigure}[b] {0.32\linewidth}
\centering
\includegraphics[width=\linewidth]{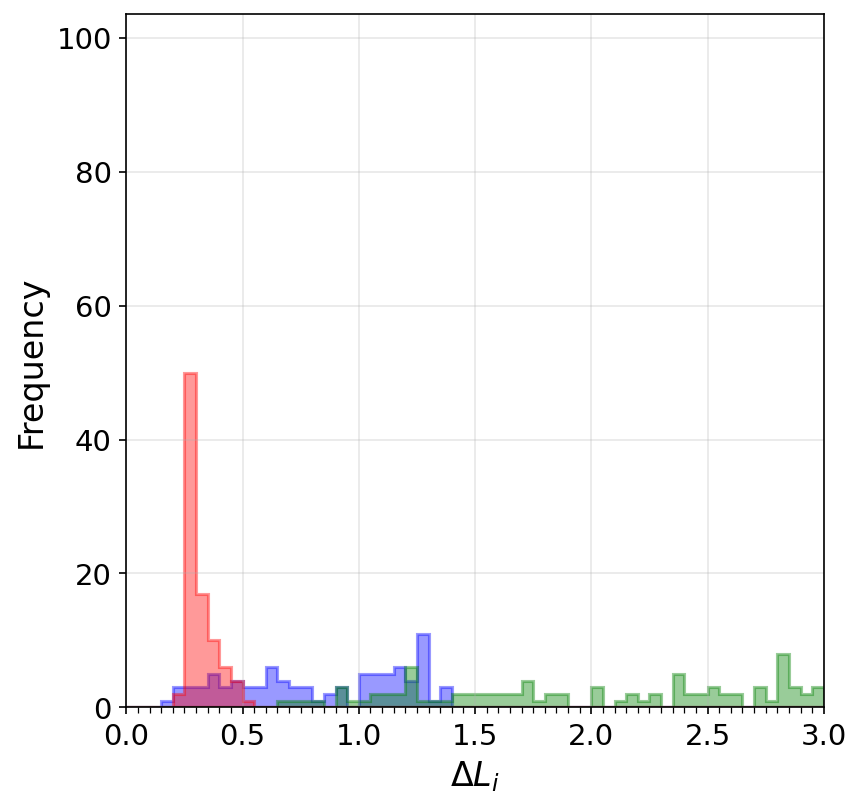}
\caption{\label{fig:cellular_ML_AL} Cellular dataset}
\end{subfigure}

\caption{\label{fig:AL_ML} Distribution of the normalized difference  of the average lifespan measure ($\Delta L_i$, see Eq.~(\ref{Average Lifespan of Generators Measure})) over 90 testing datasets.}
\end{figure}

The distributions of the normalized average lifespan measure $\Delta L_i$, defined in Eq.~\eqref{Average Lifespan of Generators Measure}, shown in Fig.~\ref{fig:AL_ML}, mirror to some extent the unpredictable behavior observed for $\Delta N_i$ above. In Fig.~\ref{fig:fourier_ML_AL}, the distributions for all three dimensions collapse near zero, implying the U-Net accurately reproduces the persistence structure of the original Fourier images prior to noising. We observe a slight degradation for the PuMA dataset in Fig.~\ref{fig:puma_ML_AL} but the most prominent unpredictability can be seen for the cellular dataset in Fig.~\ref{fig:cellular_ML_AL}. While the $i=2$ distribution maintains a peak near $\Delta L_i \approx 0.4$, the distributions for $i=0,1$ are broadly dispersed across the range of $[0,3]$. The heavy tails and flat profiles observed here for the cellular dataset affirm that $\Delta L_i$, like $\Delta N_i$, is not suited for robustly evaluating denoising performance across heterogeneous porous media datasets.

\subsubsection{Persistence Landscapes}\label{sec:MLPLmeasure}

\begin{figure}[H]
\begin{subfigure}[b] {0.32\linewidth}
\centering
\includegraphics[width=\linewidth]{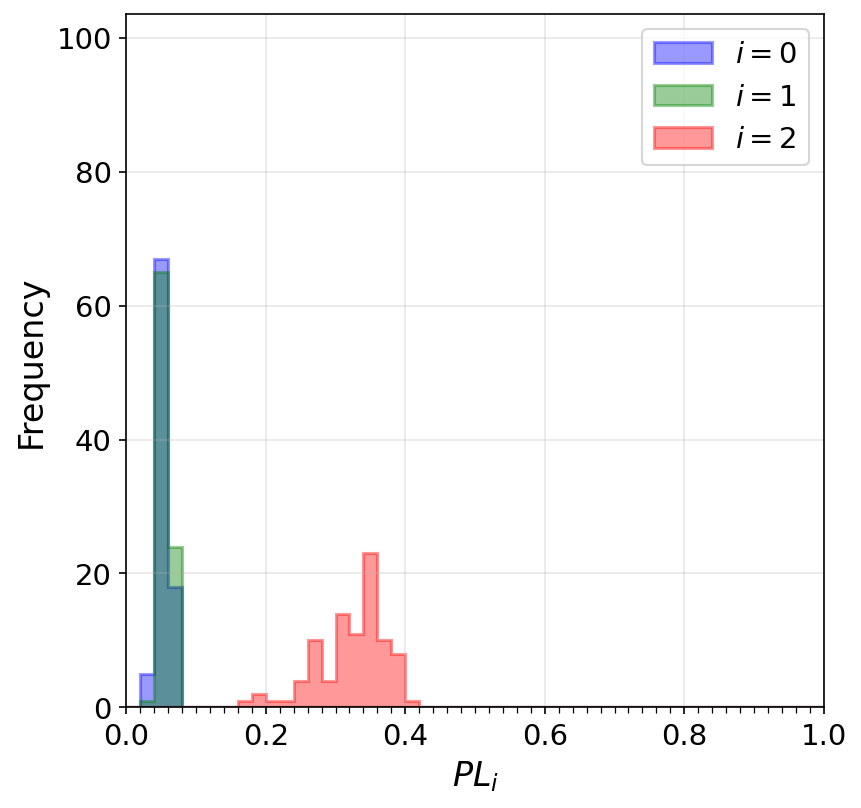}
\caption{\label{fig:fourier_ML_l2_PL} Fourier dataset}
\end{subfigure}
\hfill
\begin{subfigure}[b] {0.32\linewidth}
\centering
\includegraphics[width=\linewidth]{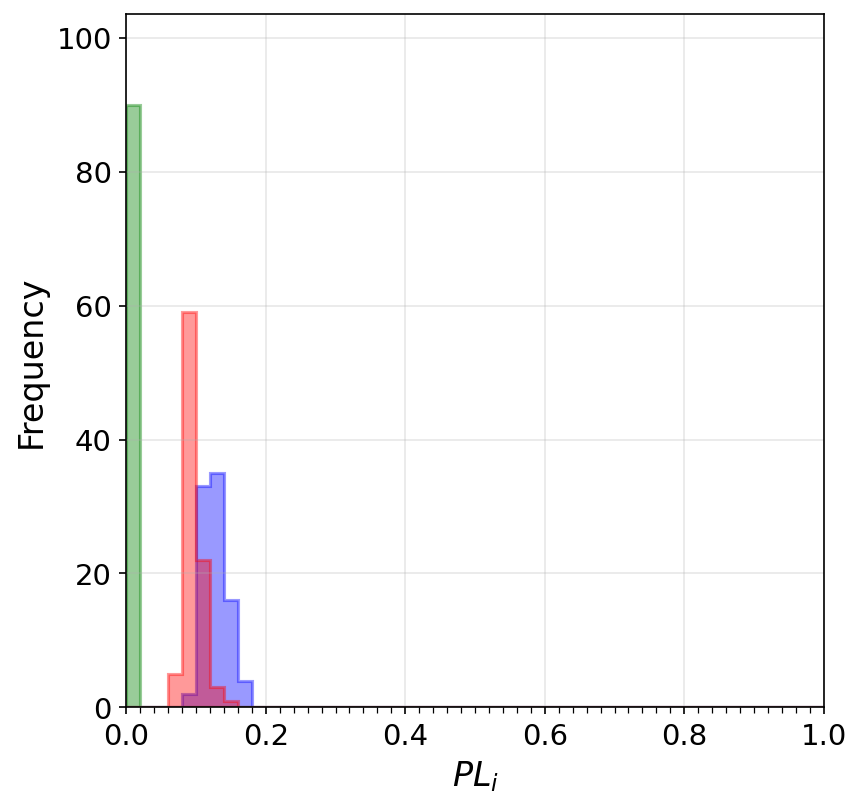}
\caption{\label{fig:puma_ML_l2_PL} PuMA dataset}
\end{subfigure}
\hfill
\begin{subfigure}[b] {0.32\linewidth}
\centering
\includegraphics[width=\linewidth]{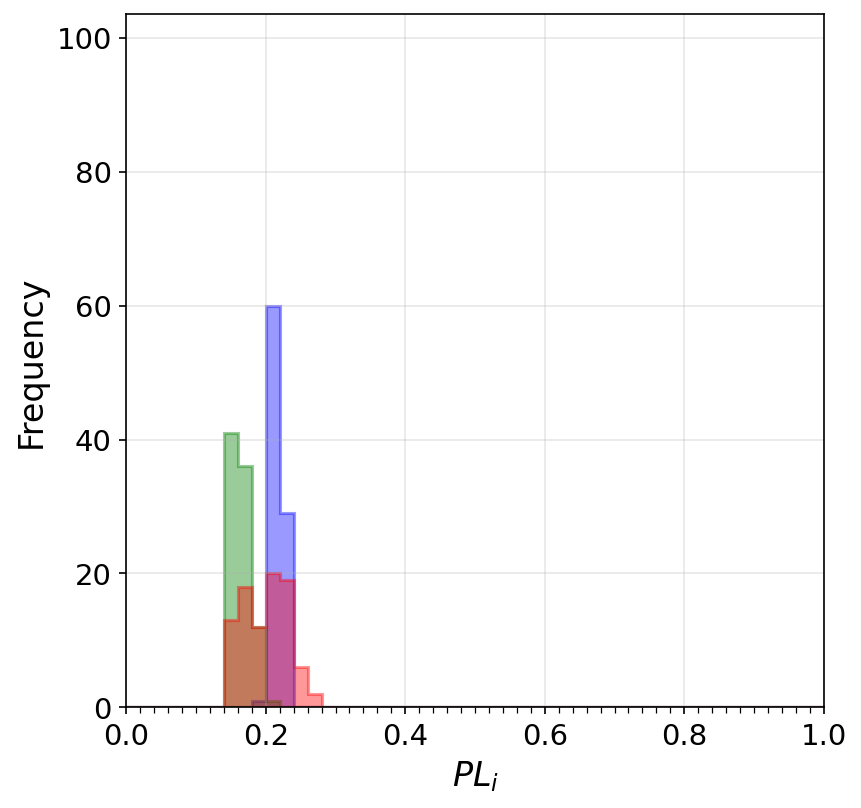}
\caption{\label{fig:cellular_ML_l2_PL} Cellular dataset}
\end{subfigure}

\caption{\label{fig:l2_PL_ML} Distribution of the normalized $L^2$ distance, $PL_{i}$ (see Sec.~\ref{sec:PLmeasure}) over 90 testing datasets.}
\end{figure}

In contrast to $\Delta N_i$ and $\Delta L_i$, the Persistence Landscape based measure $PL_i$, defined in Eq.~\eqref{PL measure}, provides substantially more consistent and interpretable distributions across all three datasets, as seen in Fig.~\ref{fig:l2_PL_ML}. For the Fourier dataset (Fig.~\ref{fig:fourier_ML_l2_PL}), the distributions of $PL_i$ for $i = 0$ and $i = 1$ are sharply concentrated near zero, while the $i = 2$ distribution is shifted to a slightly higher range ($\approx 0.2$--$0.4$) but remains compact and unimodal. The measure similarly yields tight distributions for the PuMA dataset as evidenced in Fig.~\ref{fig:puma_ML_l2_PL} with peaks below $PL_i = 0.2$ and notably less pronounced spread as compared to the persistence statistic measures described in Sections~\ref{sec:MLnumbergenerators} and~\ref{sec:MLaveragelifespan} above. The cellular dataset (Fig.~\ref{fig:cellular_ML_l2_PL}) also demonstrates well-concentrated distributions, with all three dimensions peaking around $PL_i \approx 0.2$ and exhibiting minimal spread beyond $PL_i \approx0.3$. This consistency across datasets reflects the key advantage of persistence landscapes as a functional summary: by encoding the full persistence diagram into a continuous $L^2$ function rather than a single scalar, $PL_i$ smooths over discrete topological fluctuations and provides a more stable aggregate measure of the distance between the reconstructed and ground-truth topological structures \cite{bubenik2015statistical}. Furthermore, the consistent compactness of these distributions supports the conclusion that the U-Net is able to preserve the broad persistence landscape structure across diverse porous media geometries.

\subsubsection{Persistence Images}\label{sec:MLpersistenceimage}

\begin{figure}[H]
\begin{subfigure}[b] {0.32\linewidth}
\centering
\includegraphics[width=\linewidth]{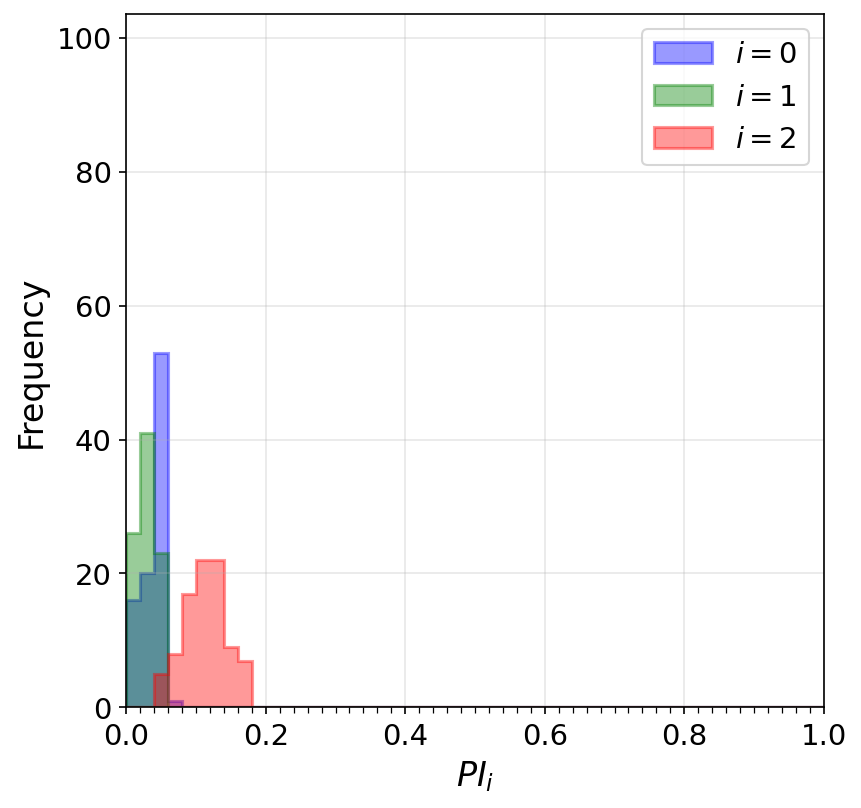}
\caption{\label{fig:fourier_ML_l2_PI} Fourier dataset}
\end{subfigure}
\hfill
\begin{subfigure}[b] {0.32\linewidth}
\centering
\includegraphics[width=\linewidth]{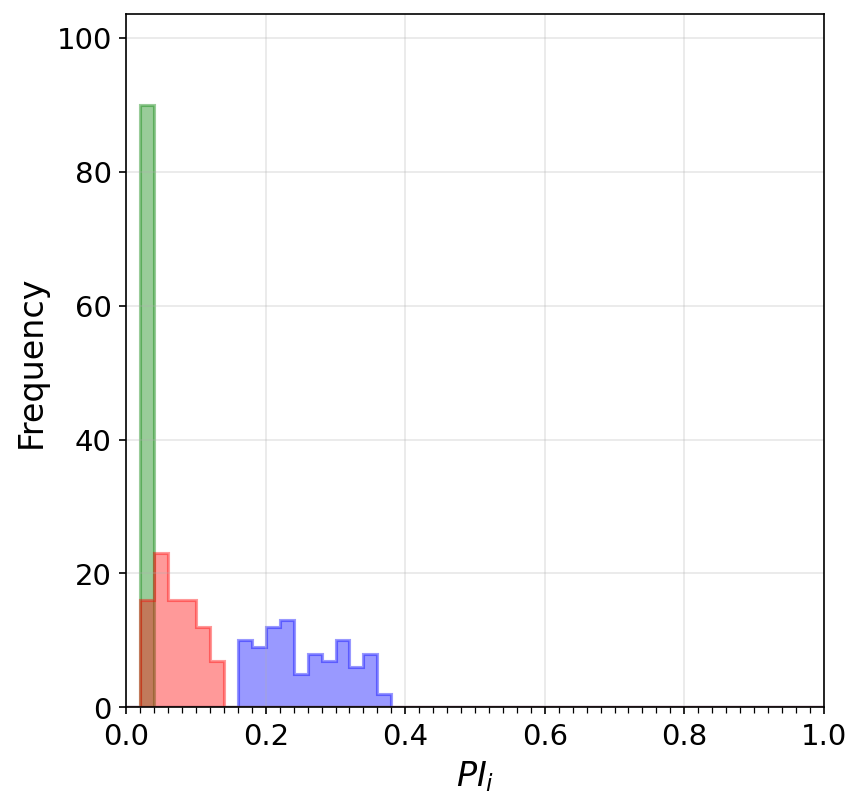}
\caption{\label{fig:puma_ML_l2_PI} PuMA dataset}
\end{subfigure}
\hfill
\begin{subfigure}[b] {0.32\linewidth}
\centering
\includegraphics[width=\linewidth]{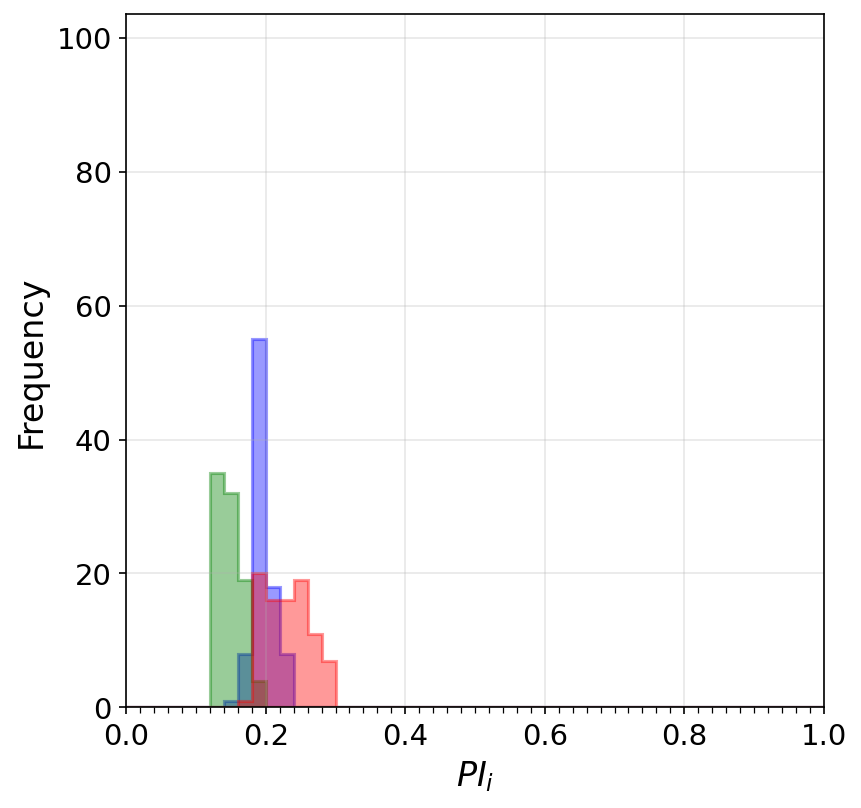}
\caption{\label{fig:cellular_ML_l2_PI} Cellular dataset}
\end{subfigure}

\caption{\label{fig:l2_PI_ML} Distribution of the normalized $L^2$ distance $PI_{i}$ (see Sec.~\ref{sec:persistenceimage}) over 90 testing datasets.}
\end{figure}

Results for the persistence image based measure $PI_i$, defined in Eq.~\eqref{PI measure}, are presented in Fig.~\ref{fig:l2_PI_ML}, and demonstrate robustness comparable to the $PL_i$ measure just discussed. For the Fourier dataset (Fig.~\ref{fig:fourier_ML_l2_PI}), the distributions for $i = 0$ and $i = 1$ are sharply peaked near zero, with the $i = 2$ distribution showing a moderate spread up to $PI_i \approx 0.2$, consistent with the patterns observed for $PL_i$. The PuMA dataset shows greater variability than was seen for $PL_i$, as seen in Fig.~\ref{fig:puma_ML_l2_PI}, with the $i = 0$ distribution exhibiting a flatter profile extending to $PI_i \approx 0.4$; this is likely attributable to the sensitivity of the persistence image construction to the choice of kernel bandwidth and weighting function, which can amplify differences near the diagonal of the persistence diagram \cite{adams2017}. Nevertheless, the $i = 0, 1, 2$ distributions remain concentrated. The cellular dataset (Fig.~\ref{fig:cellular_ML_l2_PI}) yields distributions closely aligned with those of $PL_i$, with all three dimensions concentrated below $PI_i \approx 0.4$ and clear unimodal peaks near $0.2$. The overall consistency of $PI_i$ distributions across datasets highlights the effectiveness of the persistent image representation in capturing topological differences in a noise-robust manner. Taken together, the results indicate that vectorized, $L^2$-based topological descriptors are more robust indicators of overall denoising performance as compared to measures based on persistence statistics such as the number of generators and average lifespan.

\section{Discussion\label{sec:discussion}}

Since our difference measures on the denoised and original datasets are normalized, we are able to compare their robustness to the noising/denoising process. We first discuss our findings for the Gaussian denoising method presented in Sec. \ref{sec:gaussian}, then turn our attention to the ML-based results of Sec. \ref{sec:ML}.

For the Gaussian denoising, with the partial exception of the average lifespan measure, the measures generally exhibit consistent features, notably: there is an initial error between the noisy and original datasets; the error is reduced by increasing the denoising parameter $\sigma$ up to some ``optimal'' value; and, as $\sigma$ is  increased, the   normalized error  limits to 1. The ``optimal'' $\sigma$ value  typically  occurs near $ \sigma =0.5$, which also corresponds to the value that minimizes the $ L^\infty$ distance $\|f_o-f_d\|_{\infty}$ between the original and denoised grayscale functions for the datasets. 
 Furthermore, for some measures, specifically the bottleneck distance, Wasserstein distance, persistence landscape and persistence image measures (see Figs.~\ref{fig:Bottleneck}, \ref{fig:Wasserstein}, \ref{fig:L2_PL} and \ref{fig:L2_PI}) we are able to minimize our error across all relevant homology dimensions ($i=0,1,2$)  
 simultaneously with approximately the same $\sigma$-value. The other two measures are less reliable in this respect: the number of generators measure shows good results for the Fourier and PuMA datasets, but for the Cellular dataset the minimizing values of $\sigma$ for all relevant homology dimensions are not the same and cannot be minimized simultaneously (Fig.~\ref{fig:Num_Generators}). The average lifespan measure results (Fig.~\ref{fig:Average_Lifespan}) are much noisier, but generally suggest that reasonable results should be obtained for the Fourier and PuMA datasets for $\sigma\approx 0.5$, but not for the Cellular dataset. The observation that the measures limit to 1 as the smoothing parameter $\sigma\to\infty$ is as expected: for large $\sigma$ the denoising method is over-smoothing the data to the extent that it loses all essential features. 

 We observe more nuanced and dataset-dependent results for the ML-based denoising approach. Rather than a single tunable parameter $\sigma$ that controls the denoising strength, the U-Net has many pre-tuned parameters that produce a fixed denoised output for each input. Consequently, evaluating each measure involves characterizing the distribution of the measure across the $90$ test images for each dataset. The vectorized, $L^2$-based measures --- $PL_i$ and $PI_i$ --- demonstrate consistent performance across all three datasets and homological dimensions, with compact unimodal distributions peaking within the range $0-0.4$ (see Figs.~\ref{fig:l2_PL_ML} and \ref{fig:l2_PI_ML}). These results mirror the robustness observed for $WD_i$, $BD_i$, $PL_i$, and $PI_i$ measures under Gaussian denoising at optimal $\sigma$. The persistence statistic-based measures, by contrast, are observed to be less consistent and dataset-dependent. While $\Delta N_i$ performs consistently for the Fourier dataset, with distributions concentrated near zero across all dimensions, it exhibits heavy tails and broadly dispersed distributions for the more complex PuMA and cellular datasets (see Fig.~\ref{fig:NG_ML}), echoing the dataset-dependent unreliability observed under Gaussian convolution. We observe a similar pattern for $\Delta L_i$: while well-behaved for the Fourier and PuMA datasets, we see broadly dispersed distributions for the cellular dataset (see Fig.~\ref{fig:AL_ML}). These results render $\Delta N_i$ and $\Delta L_i$ as unreliable indicators of denoising quality across diverse porous media geometries. Altogether, we find that the measure robustness hierarchy identified under Gaussian denoising --- with $L^2$ and distance-based measures outperforming scalar persistence statistics --- carries over to the ML setting, in spite of the ML framework introducing additional dataset sensitivity as compared to the parameter-controlled Gaussian framework.

In this study we used synthetic data, which offer the significant advantage that the ground truth is known prior to the addition of noise, and can be used to evaluate our measures.
Ideally, we would be able to denoise an experimental dataset with an optimal value of $\sigma$, also determined experimentally; as yet this challenge is unresolved. 
Nonetheless, we can still make some useful observations based on our results here. A robust measure should have certain properties, including: 
\begin{enumerate}[(i)]
\item the measure should not be unduly sensitive to the level of denoising; and 
\item for a reasonable range of the denoising parameter, it is decently close to the optimum value.  
\end{enumerate}
Certain measures demonstrated a greater degree of robustness with respect to these criteria, in particular the bottleneck and Wasserstein distances, and the persistence landscape measure. 
 The performance of persistence images is intermediate, and the remaining measures -- those pertaining to average lifespan of topological features and number of topological generators -- appear to be less robust. The average lifespan measure (at least for the PuMA and Cellular datasets) appears to be particularly unreliable. We hypothesize that this is because denoising, on one hand, shortens the lifespan of all of the intervals (acting to lower average lifespan); but on the other hand it also causes already short intervals to disappear (raising average lifespan). As such, the measure is rather sensitive to the level of denoising, leading to unreliable results. 

The observation that the measure based on persistence landscapes is more robust than that based on persistence images invites further comment. First, recall that we used only the first 100 persistence landscapes due to computational constraints; with thousands to millions of persistence intervals, we did not compute all landscapes. This restriction to the 100 most prominent landscapes in itself has a denoising effect. Second, we note that one of the computational parameters in the persistence images is the choice of weighting function $\omega_{\theta,m}$ to the diagonal, see \eqref{eq:WeightFunction}. For PH generated from $\alpha$ complexes, on a point cloud sampled from a $d$ dimensional probability measure, a more stable result is obtained by weighting points by choosing the weighting parameter $m \geq d$ \cite{divol2019choice}.    However with sub/super-levelset PH, there is no such heuristic that we are aware of. We used a weighting of $m=1$, but we speculate that a larger power might help to diminish the effect of the small intervals, and make persistence images more robust to the noising/denoising process.  

As with any study, ours has its limitations: the noise we used was spatially uncorrelated, i.i.d. from a Gaussian distribution with a fixed standard deviation; and our denoising process was a simple Gaussian  with varying bandwidth. Changing any of these aspects of the study could lead to different results and may merit further investigation. For example, one could employ a denoising technique that is optimized for edge or texture preservation, but one would first need to decide which feature to optimize for. We have demonstrated that our approach is able to recover the sub/super-levelset PH as seen through a variety of measures. However, the noising process may produce an exceedingly large number of small generators, which makes certain measures of PH unreliable, and the proper amount of denoising uncertain. Overall, our recommendation is that in applications with significant noise, denoising should be undertaken with caution. Persistent statistics should be viewed with skepticism and efforts should be taken to discount the effect of small intervals.

We hope this study also helps to motivate the development of theory to better describe the numerics behind our measures, which is essential for understanding their behaviors -- something that has important implications when it comes to the explainability of machine learning pipelines used to make predictions based on denoised persistence information.

\section*{Acknowledgements}
\noindent The authors acknowledge funding from NSF DMR 2410985, DMS 2201627, ACS PRF 69100-ND9, and NJIT GHAIRI Seed Grant Program. 

\bibliographystyle{elsarticle-num}
\bibliography{references}

\end{document}